\documentclass[longauth]{aa}

\usepackage{graphicx}
\usepackage{natbib}
\usepackage{amsmath}
\usepackage{color}
\usepackage{hyperref}
\bibpunct{(}{)}{;}{a}{}{,} %%natbib format for A&A and ApJ

\newcommand{\kms}{\,km\,s$^{-1}$}
\newcommand{\dm}{$\Delta m_{15} (B)$}
\newcommand{\sbv}{$s_{\mathrm{BV}}$}
\newcommand{\um}{$\mu$m}

\newcommand{\ci}{\ion{C}{i}}
\newcommand{\cii}{\ion{C}{ii}}
\newcommand{\ciii}{\ion{C}{iii}}
\newcommand{\oi}{\ion{O}{i}}
\newcommand{\siii}{\ion{Si}{ii}}
\newcommand{\siiii}{\ion{Si}{iii}}
\newcommand{\mgii}{\ion{Mg}{ii}}
\newcommand{\caii}{\ion{Ca}{ii}}
\newcommand{\tiii}{\ion{Ti}{ii}}
\newcommand{\feii}{\ion{Fe}{ii}}
\newcommand{\feiii}{\ion{Fe}{iii}}
\newcommand{\coii}{\ion{Co}{ii}}
\newcommand{\niii}{\ion{Ni}{ii}}
\newcommand{\lam}{$\lambda$}
\newcommand{\nirad}{$^{56}$Ni}
\newcommand{\nista}{$^{58}$Ni}

\begin{document}

\title{Strong near-infrared carbon in\\the Type Ia supernova iPTF13ebh
  \thanks{This paper includes data gathered with the 6.5-meter
    Magellan Telescopes located at Las Campanas Observatory, Chile.}}

\author{
{E.~Y.~Hsiao}\inst{\ref{aar},\ref{lco}}
\and {C.~R.~Burns}\inst{\ref{sbs}}
\and {C.~Contreras}\inst{\ref{lco},\ref{aar}}
\and {P.~H\"{o}flich}\inst{\ref{fsu}}
\and {D.~Sand}\inst{\ref{ttu}}
\and {G.~H.~Marion}\inst{\ref{uta},\ref{cfa}}
\and {M.~M.~Phillips}\inst{\ref{lco}}
\and {M.~Stritzinger}\inst{\ref{aar}}
\and {S.~Gonz\'alez-Gait\'an}\inst{\ref{mia},\ref{cal}}
\and {R.~E.~Mason}\inst{\ref{gno}}
\and {G.~Folatelli}\inst{\ref{ipm}}
\and {E.~Parent}\inst{\ref{bus}}
\and {C.~Gall}\inst{\ref{aar}}
\and {R.~Amanullah}\inst{\ref{okc}}
\and {G.~C.~Anupama}\inst{\ref{iia}}
\and {I.~Arcavi}\inst{\ref{lcg},\ref{ucs}}
\and {D.~P.~K.~Banerjee}\inst{\ref{nav}}
\and {Y.~Beletsky}\inst{\ref{lco}}
\and {G.~A.~Blanc}\inst{\ref{sbs},\ref{cal}}
\and {J.~S.~Bloom}\inst{\ref{ucb}}
\and {P.~J.~Brown}\inst{\ref{a&m}}
\and {A.~Campillay}\inst{\ref{lco}}
\and {Y.~Cao}\inst{\ref{cit}}
\and {A.~De~Cia}\inst{\ref{wis}}
\and {T.~Diamond}\inst{\ref{fsu}}
\and {W.~L.~Freedman}\inst{\ref{sbs}}
\and {C.~Gonzalez}\inst{\ref{lco}}
\and {A.~Goobar}\inst{\ref{okc}}
\and {S.~Holmbo}\inst{\ref{aar}}
\and {D.~A.~Howell}\inst{\ref{lcg},\ref{ucs}}
\and {J.~Johansson}\inst{\ref{okc}}
\and {M.~M.~Kasliwal}\inst{\ref{sbs}}
\and {R.~P.~Kirshner}\inst{\ref{cfa}}
\and {K.~Krisciunas}\inst{\ref{a&m}}
\and {S.~R.~Kulkarni}\inst{\ref{cit}}
\and {K.~Maguire}\inst{\ref{eso}}
\and {P.~A.~Milne}\inst{\ref{uas}}
\and {N.~Morrell}\inst{\ref{lco}}
\and {P.~E.~Nugent}\inst{\ref{ccc},\ref{ucb}}
\and {E.~O.~Ofek}\inst{\ref{wis}}
\and {D.~Osip}\inst{\ref{lco}}
\and {P.~Palunas}\inst{\ref{lco}}
\and {D.~A.~Perley}\inst{\ref{cit}}
\and {S.~E.~Persson}\inst{\ref{sbs}}
\and {A.~L.~Piro}\inst{\ref{sbs}}
\and {M.~Rabus}\inst{\ref{cat}}
\and {M.~Roth}\inst{\ref{lco}}
\and {J.~M.~Schiefelbein}\inst{\ref{a&m}}
\and {S.~Srivastav}\inst{\ref{iia}}
\and {M.~Sullivan}\inst{\ref{ush}}
\and {N.~B.~Suntzeff}\inst{\ref{a&m}}
\and {J.~Surace}\inst{\ref{ssc}}
\and {P.~R.~Wo\'zniak}\inst{\ref{lal}}
\and {O.~Yaron}\inst{\ref{wis}}
}

\institute{
Department of Physics and Astronomy, Aarhus University, Ny Munkegade 120, DK-8000 Aarhus C, Denmark \email{hsiao@phys.au.dk}\label{aar}
\and Carnegie Observatories, Las Campanas Observatory, Colina El Pino, Casilla 601, Chile\label{lco}
\and Carnegie Observatories, 813 Santa Barbara St, Pasadena, CA 91101, USA\label{sbs}
\and Florida State University, Tallahassee, FL 32306, USA\label{fsu}
\and Physics Department, Texas Tech University, Lubbock, TX 79409, USA\label{ttu}
\and University of Texas at Austin, 1 University Station C1400, Austin, TX, 78712-0259, USA\label{uta}
\and Harvard-Smithsonian Center for Astrophysics, 60 Garden Street, Cambridge, MA 02138, USA\label{cfa}
\and Millennium Institute of Astrophysics, Casilla 36-D, Santiago, Chile\label{mia}
\and Departamento de Astronom\'ia, Universidad de Chile, Casilla 36-D, Santiago, Chile\label{cal}
\and Gemini Observatory, Northern Operations Center, Hilo, HI 96720, USA\label{gno}
\and Institute for the Physics and Mathematics of the Universe (IPMU), University of Tokyo, 5-1-5 Kashiwanoha, Kashiwa, Chiba 277-8583, Japan\label{ipm}
\and Department of Physics, Bishop’s University, Sherbrooke, Quebec, J1M 1Z7 Canada\label{bus}
\and The Oskar Klein Centre, Physics Department, Stockholm University, Albanova University Center, SE 106 91 Stockholm, Sweden\label{okc}
\and Indian Institute of Astrophysics, Koramangala, Bangalore 560034, India\label{iia}
\and Las Cumbres Observatory Global Telescope Network, Goleta, CA 93117, USA\label{lcg}
\and Physics Department, University of California, Santa Barbara, CA 93106, USA\label{ucs}
\and Astronomy and Astrophysics Division, Physical Research Laboratory, Navrangapura, Ahmedabad - 380009, Gujarat, India\label{nav}
\and Department of Astronomy, University of California, Berkeley, CA, 94720-3411, USA\label{ucb}
\and Department of Physics and Astronomy, Texas A\&M University, College Station, TX 77843, USA\label{a&m}
\and Cahill Center for Astrophysics, California Institute of Technology, Pasadena, CA 91125, USA\label{cit}
\and Computational Cosmology Center, Computational Research Division, Lawrence Berkeley National Laboratory, 1 Cyclotron Road MS 50B-4206, Berkeley, CA 94611, USA\label{ccc}
\and European Southern Observatory for Astronomical Research in the Southern Hemisphere (ESO), Karl-Schwarzschild-Str. 2, 85748 Garching b. M\"{u}nchen, Germany\label{eso}
\and University of Arizona, Steward Observatory, 933 North Cherry Avenue, Tucson, AZ 85719, USA\label{uas}
\and Department of Particle Physics and Astrophysics, Weizmann Institute of Science, Rehovot 76100, Israel\label{wis}
\and Instituto de Astrof\'isica, Facultad de F\'isica, Pontificia Universidad Cat\'olica de Chile, Chile\label{cat}
\and School of Physics and Astronomy, University of Southampton, Southampton SO17 1BJ, UK\label{ush}
\and Spitzer Science Center, MS 314-6, California Institute of Technology, Pasadena, CA 91125, USA\label{ssc}
\and Space \& Remote Sensing, MS B244, Los Alamos National Laboratory, Los Alamos, NM 87545, USA\label{lal}
}

\clearpage

%%%%%%%%%%%%%%
%% Abstract %%
%%%%%%%%%%%%%%

\abstract{We present near-infrared (NIR) time-series spectroscopy, as
  well as complementary ultraviolet (UV), optical, and NIR data, of
  the Type Ia supernova (SN~Ia) iPTF13ebh, which was discovered within
  two days from the estimated time of explosion.  The first NIR
  spectrum was taken merely $2.3$ days after explosion and may be the
  earliest NIR spectrum yet obtained of a SN~Ia.  The most striking
  features in the spectrum are several NIR \ci\ lines, and the
  \ci\ \lam1.0693 \um\ line is the strongest ever observed in a SN~Ia.
  Interestingly, no strong optical \cii\ counterparts were found, even
  though the optical spectroscopic time series began early and is
  densely-cadenced.  Except at the very early epochs, within a few
  days from the time of explosion, we show that the strong NIR
  \ci\ compared to the weaker optical \cii\ appears to be general in
  SNe~Ia.  iPTF13ebh is a fast decliner with \dm$=1.79\pm0.01$, and
  its absolute magnitude obeys the linear part of the width-luminosity
  relation.  It is therefore categorized as a ``transitional'' event,
  on the fast-declining end of normal SNe~Ia as opposed to
  subluminous/91bg-like objects.  iPTF13ebh shows NIR spectroscopic
  properties that are distinct from both the normal and
  subluminous/91bg-like classes, bridging the observed characteristics
  of the two classes.  These NIR observations suggest composition and
  density of the inner core similar to that of 91bg-like events, and a
  deep reaching carbon burning layer not observed in slower declining
  SNe~Ia.  There is also a substantial difference between the
  explosion times inferred from the early-time light curve and the
  velocity evolution of the \siii\ \lam0.6355 \um\ line, implying a
  long dark phase of $\sim 4$ days.}

\titlerunning{NIR carbon of SN~Ia iPTF13ebh} 
\authorrunning{Hsiao et~al.}  

\maketitle

%%%%%%%%%%%%%%%%%%
%% Introduction %%
%%%%%%%%%%%%%%%%%%

\section{Introduction}
\label{s:intro}

%Cosmology

Type Ia supernovae (SNe~Ia), with empirically calibrated luminosities,
provide a direct measure on the expansion history of the universe and
led to the discovery of the accelerated expansion
\citep{1998AJ....116.1009R, 1999ApJ...517..565P}.  The goal of
reaching the $1-2$\% distance precision \citep{2006astro.ph..9591A} or
$\sigma_w \approx 2 \sigma_m$ \citep{2011ARNPS..61..251G} for the
next-generation SN~Ia experiments critically depends on minimizing the
systematic errors in the distance determinations.  One of the most
important is the potential evolution of the empirical calibration with
redshift.  Given the large range of look-back time considered, such an
evolution could be caused by changes in the mean host metallicity and
stellar mass with redshift \citep[e.g.,][]{2010MNRAS.406..782S}.  A
deeper understanding of the physics of SNe~Ia could help to estimate
this systematic error and perhaps even mitigate its effects.

%Explosion models and carbon

The general consensus for the origin of a SN~Ia is the thermonuclear
explosion of a carbon-oxygen white dwarf \citep{1960ApJ...132..565H}.
Since oxygen is also produced from carbon burning, carbon provides the
most direct probe of the primordial material from the progenitor.  The
quantity, distribution, and incidence of unburned carbon in SNe~Ia
provide important constraints for explosion models.  Turbulent
deflagration models predict that a large amount of unprocessed carbon
should be left over \citep{2003Sci...299...77G, 2007ApJ...668.1132R}.
Three-dimensional simulations of pure deflagrations predict a
significant amount of unburned material in the inner ejecta, in
conflict with observations \citep[e.g.,][]{2005A&A...437..983K}.  In
one-dimensional simulations, a transition to a detonation is expected
to result in complete carbon burning \citep{2002ApJ...568..791H,
  2006ApJ...645.1392M, 2009Natur.460..869K}, although there are
exceptions.  Three-dimensional effects in delayed detonation models
can create pockets of unburned material in the ejecta
\citep{2004PhRvL..92u1102G}.  In the case of pulsating delayed
detonation models \citep[PDD;][]{1996ApJ...472L..81H}, the initial
pulsation allows the outer ejecta to become loosely bound.  The low
density of this material subsequently quenches the nuclear burning.
This leaves more unburned carbon than the standard delayed detonation
models \citep{2014MNRAS.441..532D}.

%Observations of carbon

The identification of the weak absorption feature near 0.63 \um\ in
the normal SN~Ia SN~1998aq as \cii\ \lam0.6580 \um\ was first
suggested by \citet{2003AJ....126.1489B}.  \citet{2004AJ....128..387G}
then noted possible contributions of \cii\ and \ciii\ in SN~1999aa.
\citet{2007ApJ...654L..53T} presented convincing detections of the
\cii\ \lam\lam 0.4267, 0.6580, and 0.7234 lines in the early optical
spectra of SN~2006D.  \cii\ \lam0.6580 \um\ has also been detected in
the luminous and slowly expanding super-Chandrasekhar candidates:
strong \cii\ lines in SN~2009dc \citep{2011MNRAS.410..585S,
  2011MNRAS.412.2735T}, and marginal detections in SN~2003fg
\citep{2006Natur.443..308H} and SN~2007if
\citep{2010ApJ...713.1073S}. In subsequent studies with larger optical
spectroscopic samples, $20-30$\%\ of the pre-maximum spectra were
found to show \cii\ signatures \citep{2011ApJ...743...27T,
  2011ApJ...732...30P, 2012ApJ...745...74F, 2012MNRAS.425.1917S}.
Several of these studies noted, however, that this fraction represents
a lower limit, as noise, line overlap, and the phase at which the
SNe~Ia were observed could affect the \cii\ detection
\citep[e.g.,][]{2007PASP..119..709B, 2012ApJ...745...74F}.

%General properties found by CII

The velocities of the detected \cii\ lines are generally low.  There
have also been hints that SNe~Ia with detected \cii\ have
preferentially bluer colors and narrower light curves
\citep{2011ApJ...743...27T, 2012ApJ...745...74F, 2012MNRAS.425.1917S,
  2014MNRAS.444.3258M}, but this preference was not clear in the
examinations of other data sets \citep[e.g.,][]{2011ApJ...732...30P,
  2012AJ....143..126B}.  \citet{2013ApJ...779...23M} also noted a
strong preference for blue ultraviolet (UV) colors in SNe~Ia with
carbon detections.  There appears to be a consensus that the mass
fraction of carbon is low \citep[e.g.,][]{2003AJ....126.1489B,
  2006ApJ...645.1392M, 2007ApJ...654L..53T, 2008ApJ...677..448T}.
SNe~Ia that display signatures of carbon tend to have lower
\siii\ velocities, while the objects without carbon span the entire
range of \siii\ velocities \citep[e.g.,][]{2012ApJ...745...74F,
  2012MNRAS.425.1917S}.  This phenomenon may be an observational bias
as the \cii\ \lam0.6580 \um\ at high velocities is shifted into the
prominent \siii\ \lam0.6355 \um\ line.  It could also be a physical
bias.  The PDD models of \citet{2014MNRAS.441..532D} predict both
strong carbon lines and relatively narrow \siii\ \lam0.6355
\um\ absorption, while their standard delayed-detonation models
predict no carbon and broad \siii\ lines.  Several characteristics of
unburned material established by the optical \cii\ lines may be
uncertain because of the observational challenges posed by capturing
an unbiased sample of these features.

%Carbon in the NIR

Near-infrared (NIR) spectroscopy of SNe~Ia has been shown to provide
links between observables and explosion physics
\citep[e.g.,][]{1998ApJ...496..908W, 2004ApJ...617.1258H,
  2006ApJ...652L.101M, 2014ApJ...792..120F, 2014arXiv1410.6759D}.
\citet{2006ApJ...645.1392M} pioneered the study of NIR carbon
features.  Examining the \ci\ lines in NIR spectra of three normal
SNe~Ia, they concluded that the abundance of unprocessed material is
low.  The early discovery of the nearby SN~2011fe offered an
unprecedented opportunity to study the NIR carbon features in detail.
With the aid of high signal-to-noise NIR spectra and the automated
spectrum synthesis code \texttt{SYNAPPS} \citep{2011PASP..123..237T},
\citet{2013ApJ...766...72H} inferred the presence of the
\ci\ \lam1.0693 \um\ line in SN~2011fe from the flattening of the
emission component of the \mgii\ \lam1.0092 \um\ P-Cygni profile.  The
NIR \ci\ line has the same velocity as the optical \cii\ \lam0.6580
\um\ line and increases in strength toward maximum light, in contrast
to the fast-fading nature of the optical \cii\ line.
\citet{2013ApJ...766...72H} suggested that this delayed onset of the
NIR \ci\ line is an ionization effect and pointed out its potential to
secure more representative properties of unburned material in SNe~Ia.

%iPTF13ebh

iPTF13ebh is a SN~Ia discovered at an exceptionally young age by the
intermediate Palomar Transient Factory (iPTF).  The first NIR spectrum
was taken merely 2.3 days past the explosion, allowing the examination
of material in the outermost ejecta.  Strong NIR \ci\ \lam1.0693
\um\ is present in the first two NIR spectra.  To our knowledge, this
marks the fourth detection of NIR \ci\ so far in a SN~Ia, along with
SNe~1999by \citep{2002ApJ...568..791H}, 2011fe
\citep{2013ApJ...766...72H}, and 2014J \citep{2015ApJ...798...39M}.
In fact, the \ci\ feature of iPTF13ebh is the strongest ever observed.
As the supernova evolved, it became clear that the time evolution of
the NIR \ci\ feature was very different from that of any other
previous detections, and that iPTF13ebh had fast-declining light
curves.  Examination of the optical spectra showed no apparent
\tiii\ feature that would qualify it as a 91bg-like object
\citep{1992AJ....104.1543F}.  The NIR primary maxima also occurred
before the $B$ maximum.  This places iPTF13ebh in the class of
``transitional'' objects, whose notable members include SNe 1986G
\citep{1987PASP...99..592P}, 2003gs \citep{2009AJ....138.1584K},
2004eo \citep{2007MNRAS.377.1531P, 2008MNRAS.386.1897M}, 2009an
\citep{2013MNRAS.430..869S}, 2011iv \citep{2012ApJ...753L...5F}, and
2012ht \citep{2014ApJ...782L..35Y}.  In this paper, we focus on the
examination of the NIR spectroscopic properties of iPTF13ebh, and
highlight how they differ from those of normal and 91bg-like objects.
These differences could shed light on the connections between normal
and 91bg-like objects.

%%%%%%%%%%%%%%%%%%
%% Observations %%
%%%%%%%%%%%%%%%%%%

\section{Observations and Reduction}
\label{s:obs}

\begin{figure}
\centering
\includegraphics[width=0.47\textwidth,clip=true]{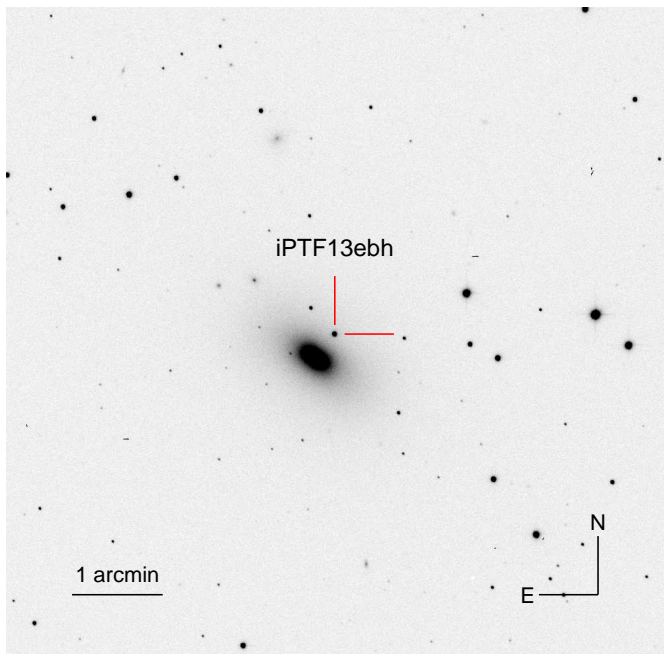}
\caption{A Swope $r$-band image of iPTF13ebh in NGC~890 taken near
  maximum light.  The cross hair marks the location of the supernova.
  The compass and the size of the field are also noted.}
\label{f:image}
\end{figure}

%Discovery

iPTF13ebh was discovered in NGC~890 by iPTF in an image taken with the
Palomar 48-inch telescope \citep[P48;][]{2008SPIE.7014E..4YR} on 2013
November 13.15 UT \citep{2013ATel.5580....1C}.  Within hours of the
discovery, the {\em Carnegie Supernova Project} (CSP) obtained a NIR
spectrum with the Folded-port Infrared Echellette
\citep[FIRE;][]{2013PASP..125..270S} mounted on the Magellan Baade
Telescope on 2013 November 14.16 UT.  Using a quick reduction
pipeline, based on the \texttt{firehose} package\footnote{Available at
  http://web.mit.edu/$^{\sim}$rsimcoe/www/FIRE/.}, a spectrum was
produced from the first two frames within 10 minutes of acquiring the
target.  The supernova was classified as a young SN\,Ia with several
strong features produced by intermediate-mass elements
\citep{2013ATel.5580....1C}.  More frames were subsequently obtained
to reach the desired signal-to-noise ratio.  An optical spectrum taken
on 2013 November 15.83 UT later confirmed the NIR classification
\citep{2013ATel.5584....1M}.  Reprocessing of the P48 images from
preceding nights showed that the supernova was visible on an image
taken on 2013 November 12.21, at $20.9\pm0.2$ mag, and not detected in
an image taken 2013 November 11.25 UT down to 21.7 mag.  Intense
follow up of iPTF13ebh was conducted in the optical and NIR as part of
the CSP, in collaboration with iPTF and the CfA Supernova Group.

\subsection{Photometric observations}

%P48 observations and reductions

The discovery and a small number of follow up images were taken with
the P48 in the $r$ band.  Host galaxy subtractions were performed
using a deep reference of NGC~890 constructed from images taken before
the explosion.  The point-spread function (PSF) was determined and
matched prior to each subtraction, and was subsequently used for PSF
photometry on the subtracted frames.  The P48 photometry is presented
in Table~\ref{t:phot_p48}.

\begin{table}
\caption{Journal of P48 photometric observations}
\label{t:phot_p48}
\centering
{\footnotesize
\begin{tabular}{cc}
\hline\hline
MJD & $r$ \\
\hline\\[-2ex]
56608.21 & 20.94 (0.21) \\
56609.15 & 18.89 (0.06) \\
56623.09 & 15.14 (0.01) \\
56624.08 & 15.15 (0.01) \\
56625.26 & 15.15 (0.01) \\
56626.14 & 15.16 (0.01) \\
56648.09 & 16.60 (0.01) \\
\hline
\end{tabular}
} \tablefoot{The MJD column lists the Modified Julian Date of each
  observation.  The P48 photometry is host galaxy subtracted.}
\end{table}

%Swope observations and reductions

Swope follow up in $uBVgri$ began on the same night that the FIRE
classification spectrum was obtained.  All images were taken with the
newly commissioned e2v CCD imager which is more efficient overall and
specifically more sensitive in the blue than the previous CCD employed
by the CSP.  The pixel size of 0\farcs435 remains the same.  The
$2\times2$ CCD array yields a field of view of approximately
30\arcmin\ on each side.  The supernova and standard fields are
normally placed at the center of one pre-selected quadrant.  The
bandpass functions with the new e2v imager have been characterized
with a spectrometer, in the same manner as described in
\citet{2010SPIE.7735E..64R} and \citet{2011AJ....142..156S}.  The
results are presented in \citet{2014SPIE.9147E..5LR}.  A Swope
$r$-band image taken near maximum light is shown in
Fig.~\ref{f:image}.  The reduction of Swope images was done as
described in \citet{2010AJ....139..519C}.  PSF photometry was
performed with respect to a local sequence of standard stars
calibrated to the \citet{1992AJ....104..340L} and
\citet{2002AJ....123.2121S} standard fields.  The standard fields were
observed on the same nights as the supernova observations over the
course of 20 photometric nights.  The Swope photometry is tabulated
in Table~\ref{t:phot_swope}.

\begin{table*}
\caption{Journal of Swope e2v photometric observations}
\label{t:phot_swope}
\centering
{\footnotesize
\begin{tabular}{ccccccc}
\hline\hline
MJD & $B$ & $V$ & $u$ & $g$ & $r$ & $i$ \\
\hline\\[-2ex]
56610.13 & 18.466 (0.024) & 18.101 (0.021) & 20.082 (0.172) & 18.230 (0.017) & 18.027 (0.020) & 18.309 (0.046) \\
56611.12 & 17.803 (0.020) & 17.501 (0.015) &       $\cdots$ & 17.595 (0.013) & 17.435 (0.018) & 17.632 (0.029) \\
56612.14 & 17.332 (0.014) & 17.134 (0.014) & 18.265 (0.063) & 17.193 (0.013) & 17.056 (0.012) & 17.166 (0.019) \\
56613.19 & 16.937 (0.019) & 16.768 (0.013) &       $\cdots$ & 16.803 (0.011) & 16.654 (0.011) & 16.740 (0.013) \\
56614.10 & 16.604 (0.009) & 16.482 (0.008) & 17.247 (0.023) & 16.497 (0.008) & 16.362 (0.007) & 16.447 (0.011) \\
56615.13 & 16.327 (0.009) & 16.184 (0.007) &       $\cdots$ & 16.192 (0.006) & 16.057 (0.006) & 16.147 (0.008) \\
56617.12 & 15.841 (0.006) & 15.720 (0.007) &       $\cdots$ & 15.711 (0.006) & 15.609 (0.008) & 15.722 (0.016) \\
56618.12 & 15.656 (0.005) & 15.560 (0.006) & 16.256 (0.009) & 15.539 (0.005) & 15.441 (0.006) & 15.603 (0.007) \\
56619.10 & 15.541 (0.006) & 15.411 (0.009) & 16.145 (0.009) & 15.395 (0.008) & 15.288 (0.009) & 15.474 (0.013) \\
56620.11 & 15.442 (0.007) & 15.352 (0.008) & 16.112 (0.012) & 15.304 (0.007) & 15.210 (0.012) & 15.486 (0.012) \\
56621.11 & 15.368 (0.007) & 15.278 (0.006) & 16.111 (0.029) & 15.296 (0.006) & 15.168 (0.006) & 15.505 (0.018) \\
56622.10 & 15.343 (0.005) & 15.216 (0.005) & 16.142 (0.009) & 15.202 (0.005) & 15.109 (0.006) & 15.423 (0.007) \\
56623.09 & 15.343 (0.006) & 15.181 (0.008) & 16.191 (0.011) & 15.189 (0.005) & 15.099 (0.009) & 15.469 (0.011) \\
56624.07 & 15.361 (0.007) & 15.167 (0.006) & 16.270 (0.013) & 15.185 (0.005) & 15.068 (0.006) & 15.480 (0.010) \\
56625.09 & 15.401 (0.006) & 15.176 (0.006) & 16.397 (0.014) & 15.211 (0.005) & 15.059 (0.007) & 15.507 (0.009) \\
56630.08 & 15.956 (0.011) & 15.388 (0.012) & 17.190 (0.026) & 15.622 (0.010) & 15.322 (0.012) & 15.841 (0.016) \\
56631.09 & 16.124 (0.007) & 15.430 (0.006) & 17.421 (0.021) & 15.703 (0.006) & 15.360 (0.007) & 15.846 (0.013) \\
56632.09 & 16.280 (0.008) & 15.509 (0.007) & 17.492 (0.015) & 15.818 (0.005) & 15.430 (0.008) & 15.883 (0.011) \\
56633.08 & 16.442 (0.011) & 15.607 (0.008) & 17.711 (0.018) & 15.953 (0.009) & 15.462 (0.010) & 15.895 (0.013) \\
56635.08 & 16.781 (0.014) & 15.794 (0.007) & 18.116 (0.030) & 16.252 (0.006) & 15.558 (0.006) & 15.895 (0.009) \\
56636.08 & 16.925 (0.014) & 15.860 (0.009) & 18.266 (0.036) & 16.392 (0.008) & 15.612 (0.009) & 15.867 (0.013) \\
56637.07 & 17.107 (0.021) & 15.958 (0.008) & 18.307 (0.047) & 16.514 (0.008) & 15.653 (0.006) & 15.859 (0.008) \\
56638.10 & 17.224 (0.017) & 16.050 (0.008) & 18.535 (0.034) & 16.678 (0.010) & 15.693 (0.008) & 15.862 (0.011) \\
56640.06 & 17.447 (0.023) & 16.206 (0.014) & 18.668 (0.047) & 16.880 (0.012) & 15.804 (0.008) & 15.886 (0.009) \\
56641.09 & 17.553 (0.031) & 16.295 (0.011) & 18.941 (0.084) & 17.012 (0.015) & 15.890 (0.007) & 15.928 (0.009) \\
56644.05 & 17.815 (0.026) & 16.557 (0.011) & 19.145 (0.092) & 17.279 (0.019) & 16.131 (0.008) & 16.050 (0.010) \\
56645.05 & 17.872 (0.019) & 16.650 (0.010) & 18.997 (0.050) & 17.354 (0.014) & 16.232 (0.008) & 16.123 (0.013) \\
56646.05 & 17.939 (0.013) & 16.694 (0.010) & 19.157 (0.030) & 17.392 (0.013) & 16.317 (0.011) & 16.220 (0.013) \\
56647.05 & 17.971 (0.014) & 16.764 (0.012) & 19.224 (0.029) & 17.440 (0.012) & 16.399 (0.011) & 16.318 (0.013) \\
56648.05 & 18.079 (0.015) & 16.830 (0.010) & 19.307 (0.035) & 17.507 (0.014) & 16.473 (0.011) & 16.397 (0.014) \\
56649.05 & 18.089 (0.015) & 16.876 (0.011) & 19.376 (0.039) & 17.529 (0.012) & 16.539 (0.012) & 16.506 (0.014) \\
56651.07 & 18.179 (0.018) & 17.017 (0.013) & 19.405 (0.038) & 17.634 (0.014) & 16.694 (0.013) & 16.634 (0.018) \\
56652.09 & 18.222 (0.023) & 17.061 (0.013) & 19.407 (0.068) & 17.700 (0.015) & 16.760 (0.017) & 16.824 (0.026) \\
56653.04 & 18.251 (0.017) & 17.121 (0.013) & 19.436 (0.028) & 17.765 (0.017) & 16.818 (0.017) & 16.767 (0.023) \\
56654.04 & 18.274 (0.020) & 17.154 (0.012) & 19.442 (0.031) & 17.761 (0.013) & 16.824 (0.012) & 16.808 (0.018) \\
56656.04 & 18.408 (0.027) & 17.273 (0.020) & 19.509 (0.053) & 17.842 (0.016) & 16.913 (0.014) & 16.911 (0.019) \\
56657.07 & 18.360 (0.016) & 17.276 (0.015) & 19.634 (0.041) & 17.840 (0.015) & 16.971 (0.015) & 16.976 (0.016) \\
56659.06 & 18.330 (0.018) &       $\cdots$ &       $\cdots$ &       $\cdots$ &       $\cdots$ &       $\cdots$ \\
\hline
\end{tabular}
} \tablefoot{The MJD column lists the Modified Julian Date of each
  observation.  The Swope photometry is without host subtraction.}
\end{table*}

%difference between P48 and Swope r

\begin{figure}
\centering
\includegraphics[width=0.47\textwidth,clip=true]{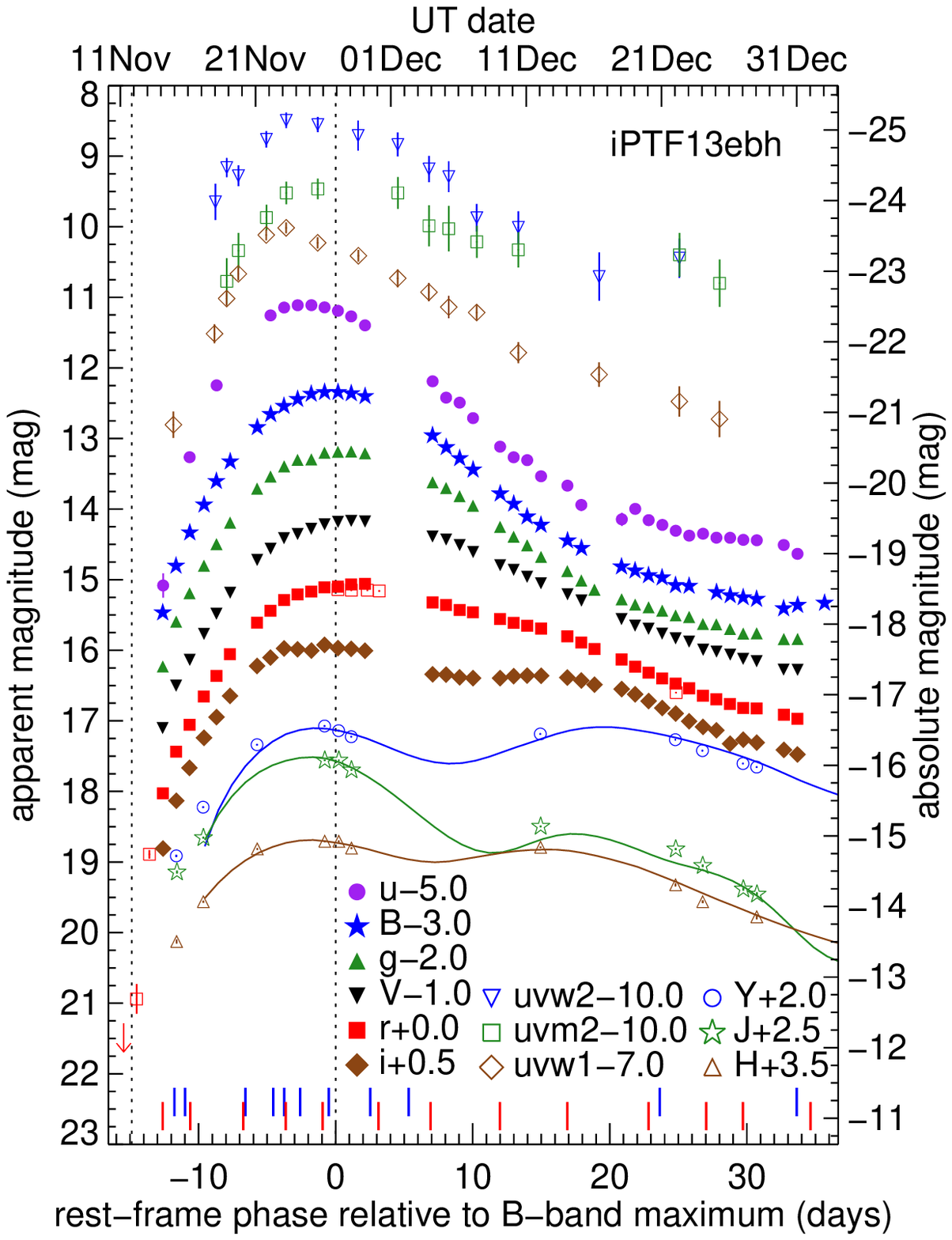}
\caption{UV, optical and NIR light curves of iPTF13ebh.  The $uvw2$,
  $uvm2$ and $uvw1$ light curves are obtained with UVOT on Swift; the
  $uBVgri$ light curves are obtained with the e2v imager on Swope; and
  the $YJH$ light curves are obtained with RetroCam on du Pont.  All
  of these are without host subtractions.  For the $r$-band light
  curve, open red square symbols represent host-subtracted $r$-band
  data from P48.  A downward arrow marks the date of a non-detection
  image from P48 down to a magnitude limit of 21.7 mag.
  \texttt{SNooPy} fits \citep{2011AJ....141...19B} are plotted for the
  more sparsely-sampled NIR light curves.  Blue and red vertical dash
  lines mark the dates when the optical and NIR spectra are taken,
  respectively.  Two black vertical dotted lines are drawn at the
  inferred time of explosion and time of $B$-band maximum.  The
  absolute magnitude is computed using a distance modulus of 33.63,
  derived from the host recession velocity of
  \citet{2006AJ....132..197W}}.
\label{f:phot}
\end{figure}

The Swope and P48 light curves are presented in the standard
\citet{1992AJ....104..340L} and \citet{2002AJ....123.2121S} systems in
Fig.~\ref{f:phot}.  Note that the Swope light curves are without host
galaxy template subtractions, and no host galaxy or Milky Way dust
extinction corrections have been applied.  From Fig.~\ref{f:image},
the site of iPTF13ebh does not appear to have substantial host galaxy
contamination.  Hence, the un-subtracted photometry should be adequate
for the determination of basic photometric parameters.  This is
corroborated by the comparison of the Swope and P48 $r$-band light
curves.  On nights when both Swope and P48 $r$-band photometry is
available, the Swope points are consistently brighter, but by only
$\sim0.01$ mag on average.

%NIR photometry

The NIR $YJH$ light curves were obtained using RetroCam, which was
moved from the 1-m Swope to the 2.5-m du Pont telescope in 2011.  The
single-chip Rockwell HAWAII-1 HgCdTe detector and 0\farcs201 pixel
size yield a field of view of 3\farcm5 on each side.  The images were
reduced in the standard manner following \citet{2010AJ....139..519C}.
As in the optical, the NIR photometry was computed differentially with
respect to a local sequence of stars.  The local sequence was then
calibrated to the \citet{1998AJ....116.2475P} system in the $JH$
bands, with the standard fields observed during six photometric
nights.  The $Y$ band was calibrated to a set of yet-to-be-published
magnitudes of the Persson standards.  The NIR light curves are also
without host galaxy template subtractions.  The RetroCam photometry is
tabulated in Table~\ref{t:phot_dupont}.

\begin{table}
\caption{Journal of du Pont RetroCam photometric observations}
\label{t:phot_dupont}
\centering
{\footnotesize
\begin{tabular}{cccc}
\hline\hline
MJD & $Y$ & $J$ & $H$ \\
\hline\\[-2ex]
56611.14 & 16.913 (0.016) & 16.643 (0.022) & 16.627 (0.028) \\
56613.12 & 16.223 (0.011) & 16.156 (0.014) & 16.063 (0.016) \\
56617.11 & 15.338 (0.009) &       $\cdots$ & 15.315 (0.011) \\
56622.10 & 15.073 (0.009) & 15.059 (0.009) & 15.206 (0.011) \\
56623.14 & 15.139 (0.009) & 15.058 (0.009) & 15.207 (0.013) \\
56624.07 & 15.225 (0.009) & 15.196 (0.009) & 15.303 (0.011) \\
56638.05 & 15.186 (0.009) & 15.996 (0.012) & 15.291 (0.012) \\
56648.03 & 15.267 (0.009) & 16.317 (0.014) & 15.824 (0.022) \\
56650.04 & 15.422 (0.009) & 16.552 (0.019) & 16.063 (0.020) \\
56653.03 & 15.605 (0.010) & 16.884 (0.023) &       $\cdots$ \\
56654.04 & 15.655 (0.011) & 16.956 (0.028) & 16.277 (0.035) \\
\hline
\end{tabular}
} \tablefoot{The MJD column lists the Modified Julian Date of each
  observation.  The RetroCam photometry is without host subtraction.}
\end{table}

%Swift UV photometry

We also present UV $uvw1$, $uvm2$, and $uvw2$ light curves obtained
with the Ultraviolet Optical Telescope
\citep[UVOT;][]{2005SSRv..120...95R} on the Swift Mission
\citep{2004ApJ...611.1005G}.  The photometry follows the method
outlined by \citet{2012ApJ...753...22B, 2014Ap&SS.tmp..301B}, which
incorporates the updated zero points and time-dependent sensitivity
corrections of \citet{2011AIPC.1358..373B}.  Note that host galaxy
templates are not yet available to remove the non-SN counts.  The UVOT
photometry is tabulated in Table~\ref{t:phot_swift}.

\begin{table}
\caption{Journal of Swift UVOT photometric observations}
\label{t:phot_swift}
\centering
{\footnotesize
\begin{tabular}{cccc}
\hline\hline
MJD & $uvw2$ & $uvm2$ & $uvw1$ \\
\hline\\[-2ex]
56610.9 &       $\cdots$ &       $\cdots$ & 19.805 (0.187) \\
56614.0 & 19.647 (0.256) &       $\cdots$ & 18.515 (0.133) \\
56614.9 & 19.162 (0.136) & 20.773 (0.326) & 18.017 (0.101) \\
56615.7 & 19.277 (0.147) & 20.339 (0.250) & 17.667 (0.092) \\
56617.8 & 18.761 (0.116) & 19.870 (0.181) & 17.117 (0.077) \\
56619.3 & 18.492 (0.110) & 19.522 (0.159) & 17.016 (0.077) \\
56621.6 & 18.551 (0.108) & 19.465 (0.146) & 17.228 (0.079) \\
56624.6 & 18.708 (0.212) &       $\cdots$ & 17.411 (0.083) \\
56627.5 & 18.835 (0.169) & 19.520 (0.225) & 17.731 (0.126) \\
56629.8 & 19.181 (0.185) & 19.987 (0.290) & 17.927 (0.130) \\
56631.3 & 19.291 (0.219) & 20.027 (0.321) & 18.138 (0.158) \\
56633.3 & 19.872 (0.194) & 20.214 (0.226) & 18.215 (0.110) \\
56636.4 & 20.009 (0.228) & 20.327 (0.249) & 18.785 (0.149) \\
56642.4 & 20.706 (0.341) &       $\cdots$ & 19.094 (0.172) \\
56648.3 & 20.445 (0.277) & 20.398 (0.309) & 19.475 (0.215) \\
56651.3 &       $\cdots$ & 20.800 (0.334) & 19.726 (0.256) \\
\hline
\end{tabular}
} \tablefoot{The MJD column lists the Modified Julian Date of each
  observation.  The UVOT photometry is without host subtraction.}
\end{table}

\subsection{Spectroscopic observations}

%optical spectra

\begin{figure}
\centering
\includegraphics[width=0.47\textwidth,clip=true]{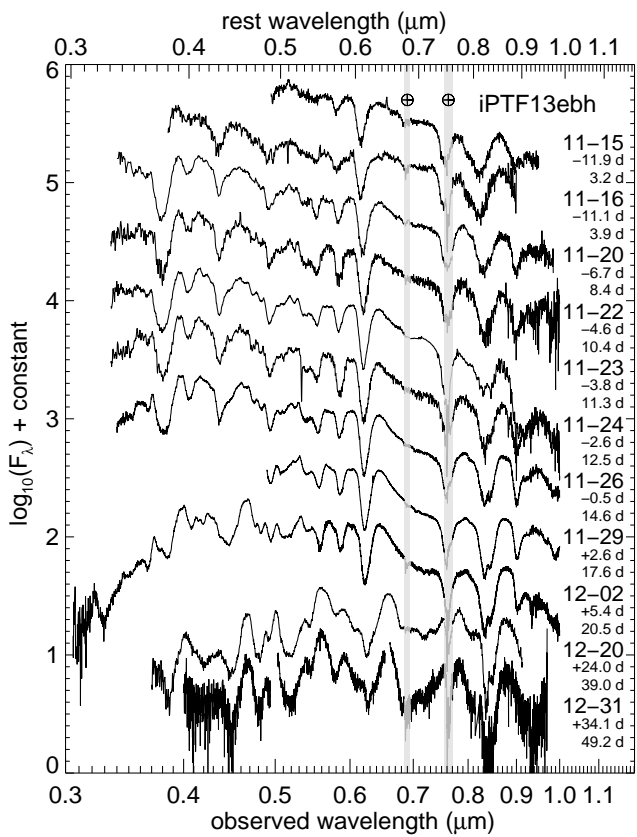}
\caption{Optical spectra of iPTF13ebh.  The UT date of observation,
  phase relative to explosion and phase relative to $B$-band maximum
  are labeled for each spectrum.  The gray vertical bands mark the
  regions of the strongest telluric absorptions.}
\label{f:optspec}
\end{figure}

Through the iPTF collaboration, low-resolution optical spectra were
obtained from a variety of instruments.  These include the
Intermediate dispersion Spectrograph and Imaging System (ISIS) on the
William Herschel Telescope (WHT), the Himalaya Faint Object
Spectrograph Camera (HFOSC) on the Himalayan Chandra Telescope (HCT),
the Dual Imaging Spectrograph (DIS) on the ARC 3.5-m telescope, the
cross-dispersed spectrograph FLOYDS on the robotic Faulkes Telescope
North (FTN), the Andalucia Faint Object Spectrograph and Camera
(ALFOSC) on the Nordic Optical Telescope (NOT), the Double Beam
Spectrograph \citep[DBSP;][]{1982PASP...94..586O} on the Palomar
200-inch telescope (P200), the Deep Imaging Multi-Object Spectrograph
\citep[DEIMOS;][]{2003SPIE.4841.1657F} on the Keck II telescope, the
Low Resolution Imaging Spectrometer
\citep[LRIS;][]{1995PASP..107..375O} on the Keck I telescope, and the
Inamori Magellan Areal Camera and Spectrograph
\citep[IMACS;][]{2011PASP..123..288D} on the Magellan telescope.  A
journal of observations is presented in Table~\ref{t:obs_opt}, and the
spectra are plotted in Fig~\ref{f:optspec}.  The spectra are made
public via the WISeREP database \citep{2012PASP..124..668Y}.

\begin{table}
\caption{Journal of optical spectroscopic observations}
\label{t:obs_opt}
\centering
{\footnotesize
\begin{tabular}{ccccccc}
\hline\hline
UT Date & MJD  & Instrument & $t_{\rm{max}}(B)$ & $t_{\rm{exp}}$\\
\hline\\[-2ex]
2013-11-15 & 56611.00 &      WHT +   ISIS & $-11.9$ &  3.2\\
2013-11-16 & 56611.79 &      HCT +  HFOSC & $-11.1$ &  3.9\\
2013-11-20 & 56616.25 &      ARC +    DIS & $ -6.7$ &  8.4\\
2013-11-22 & 56618.30 &      FTN + FLOYDS & $ -4.6$ & 10.4\\
2013-11-23 & 56619.12 &      NOT + ALFOSC & $ -3.8$ & 11.3\\
2013-11-24 & 56620.30 &      FTN + FLOYDS & $ -2.6$ & 12.5\\
2013-11-26 & 56622.41 &     P200 +   DBSP & $ -0.5$ & 14.6\\
2013-11-29 & 56625.48 &     Keck + DEIMOS & $ +2.6$ & 17.6\\
2013-12-02 & 56628.32 &     Keck +   LRIS & $ +5.4$ & 20.5\\
2013-12-20 & 56646.88 &      NOT + ALFOSC & $+24.0$ & 39.0\\
2013-12-31 & 56657.00 & Magellan +  IMACS & $+34.1$ & 49.2\\
\hline
\end{tabular}
} \tablefoot{The MJD column lists the Modified Julian Date of each
  observation.  We adopt the time of $B$ maximum of JD 2,456,623.4 and
  the time of explosion of JD 2,456,608.4 for the phase relative to
  $B$-band maximum, $t_{\rm{max}}(B)$, and phase relative to
  explosion, $t_{\rm{exp}}$, respectively.}
\end{table}

%observing strategy

\begin{figure}
\centering
\includegraphics[width=0.47\textwidth,clip=true]{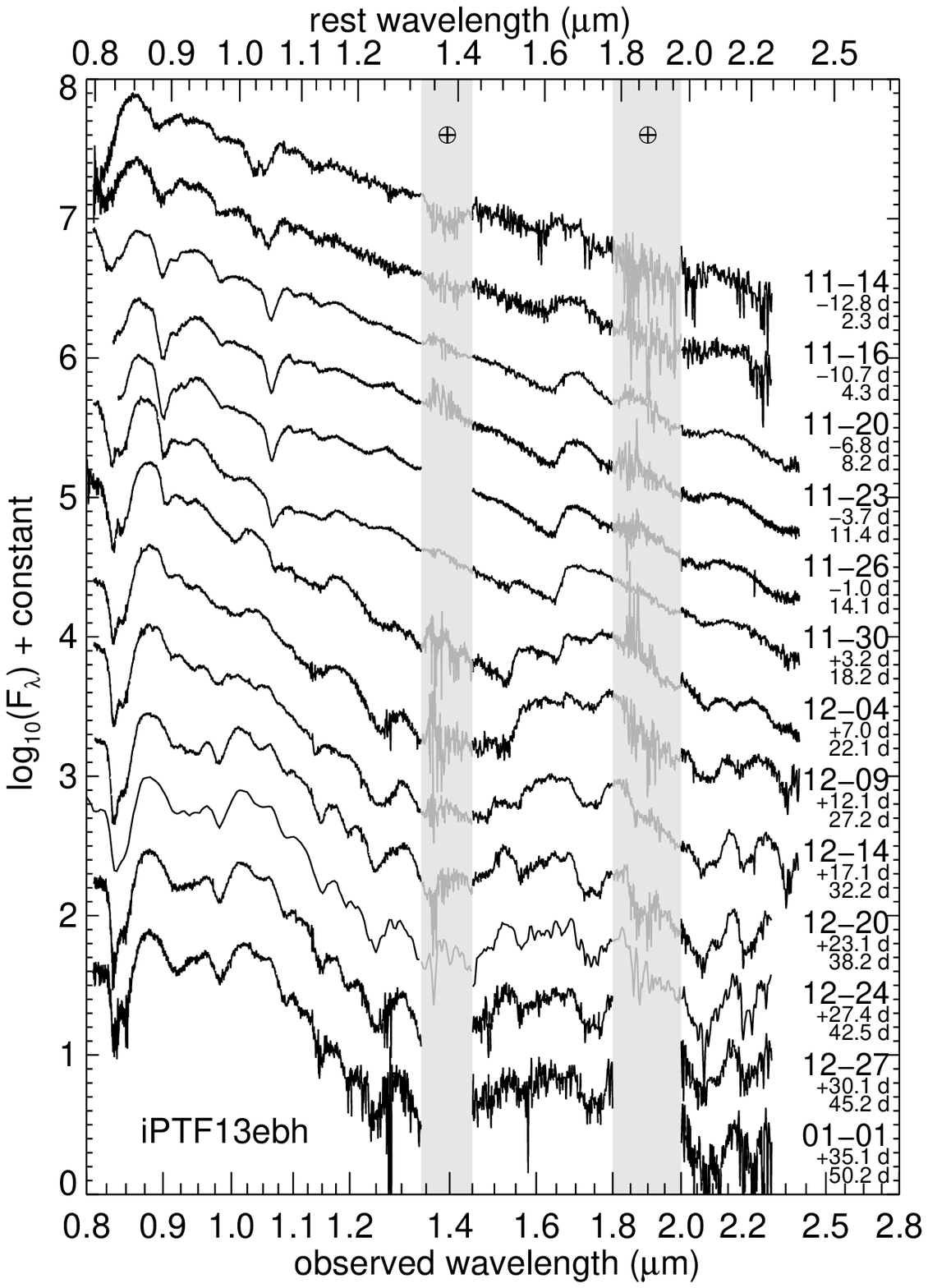}
\caption{NIR spectra of iPTF13ebh.  The UT date of observation, phase
  relative to explosion and phase relative to $B$-band maximum are
  labeled for each spectrum.  The gray vertical bands mark the regions
  of the strongest telluric absorptions.}
\label{f:nirspec}
\end{figure}

The NIR spectroscopic time series of iPTF13ebh was obtained using FIRE
\citep{2013PASP..125..270S} on the Magellan telescope, GNIRS
\citep{1998SPIE.3354..555E} on the Gemini North Telescope, and SpeX
\citep{2003PASP..115..362R} on the NASA Infrared Telescope Facility
(IRTF) as part of a joint CSP-CfA Supernova Group program to obtain
time-series NIR spectroscopy of supernovae.  The combination of
classically scheduled time on Magellan and target-of-opportunity queue
observing on Gemini North allowed spectra to be obtained in a regular
2-4 day cadence when the supernova was young.  The observing log is
presented in Table~\ref{t:obs_nir}.  The NIR spectroscopic time series
is presented in Fig.~\ref{f:nirspec}.

\begin{table}
\caption{Journal of NIR spectroscopic observations}
\label{t:obs_nir}
\centering
{\footnotesize
\begin{tabular}{ccccccc}
\hline\hline
UT Date & MJD  & Instrument & $t_{\rm{max}}(B)$ & $t_{\rm{exp}}$ & $T_{\rm{int}}$\\
\hline\\[-2ex]
2013-11-14 & 56610.14 &  FIRE & $-12.8$ d &  2.3 d & 52.8\\
2013-11-16 & 56612.17 &  FIRE & $-10.7$ d &  4.3 d & 21.1\\
2013-11-20 & 56616.10 &  FIRE & $ -6.8$ d &  8.2 d & 25.4\\
2013-11-23 & 56619.23 & GNIRS & $ -3.7$ d & 11.4 d & 40.0\\
2013-11-26 & 56621.96 & GNIRS & $ -1.0$ d & 14.1 d & 32.0\\
2013-11-30 & 56626.08 &  FIRE & $ +3.2$ d & 18.2 d & 19.0\\
2013-12-04 & 56630.43 & GNIRS & $ +7.4$ d & 22.7 d & 30.0\\
2013-12-09 & 56635.06 &  FIRE & $+12.1$ d & 27.2 d & 25.4\\
2013-12-14 & 56640.05 &  FIRE & $+17.1$ d & 32.2 d & 16.9\\
2013-12-20 & 56646.05 &  FIRE & $+23.1$ d & 38.2 d & 16.9\\  
2013-12-24 & 56650.32 &  SpeX & $+27.4$ d & 42.5 d & 30.0\\  
2013-12-27 & 56653.03 &  FIRE & $+30.1$ d & 45.2 d & 16.9\\
2014-01-01 & 56658.03 &  FIRE & $+35.1$ d & 50.2 d & 16.9\\
\hline
\end{tabular}
} \tablefoot{The MJD column lists the Modified Julian Date of each
  observation.  We adopt the time of $B$ maximum of JD 2,456,623.4 and
  the time of explosion of JD 2,456,608.4 for the phase relative to
  $B$-band maximum, $t_{\rm{max}}(B)$, and phase relative to
  explosion, $t_{\rm{exp}}$, respectively.  $T_{\rm{int}}$ represents
  the total on-target integration time in minutes.}
\end{table}

%FIRE observations

The FIRE spectra were obtained in the high-throughput prism mode with
a 0\farcs6 slit.  This configuration yields a continuous wavelength
coverage from 0.8 to 2.5 \um\ with a resolution of $R\sim500$ in the
$J$ band.  When acquiring the supernova, the slit was oriented along
the parallactic angle to minimize the effect of differential
refraction \citep{1982PASP...94..715F}.  At each epoch, several frames
were obtained using the conventional ABBA ``nod-along-the-slit''
technique and the ``sampling-up-the-ramp'' readout mode.  The
per-frame exposure time was between 95.1 and 158.5 seconds depending
on the brightness of the supernova.  These exposure times were chosen
such that an adequate signal was obtained in each frame without
saturating the bright sky lines in the $K$ band.  At each epoch, an
A0V star was observed close to the science observations in time,
angular distance and air mass for telluric correction, as per the
method described in \citet{2003PASP..115..389V}.

%FIRE reductions

The data were reduced using the IDL pipeline \texttt{firehose},
specifically designed for the reduction of FIRE data.  The pipeline
performed steps of flat fielding, wavelength calibration, sky
subtraction, spectral tracing and extraction.  The sky flux was
modeled using off-source pixels as described by
\citet{2003PASP..115..688K} and subtracted from each frame.  The
spectral extraction was then performed using the optimal technique
\citep{1986PASP...98..609H}, a weighting scheme that delivers the
maximum signal-to-noise ratio while preserving spectrophotometric
accuracy.  Individual spectra were then combined with sigma clipping to
reject spurious pixels.  Corrections for telluric absorption were 
performed using the IDL tool \texttt{xtellcor} developed by
\citet{2003PASP..115..389V}.  To construct a telluric correction
spectrum free of stellar absorption features, a model spectrum of Vega
was used to match and remove the hydrogen lines of the Paschen and
Brackett series from the A0V telluric standard.  The resulting
telluric correction spectrum was also used for flux calibration.

%GNIRS observations

The GNIRS spectra were observed in the cross-dispersed mode, in
combination with the short-wavelength camera, a 32 lines per mm
grating, and 0$\farcs$675 slit.  This configuration allows for a wide
continuous wavelength coverage from 0.8 to 2.5 \um, divided over six
orders and yields a resolution of $R\sim1000$.  The observing setup
was similar to that described for FIRE observations.  Because of the
higher resolution for GNIRS, higher per-frame exposure times, between
240 and 300 seconds, were chosen.  The slit was positioned at the
parallactic angle at the beginning of each observation.  An A0V star
was also observed after each set of science observations for telluric
and flux calibration.

%GNIRS reduction

The GNIRS data were calibrated and reduced using the \texttt{XDGNIRS}
pipeline, specifically developed for the reduction of GNIRS
cross-dispersed data.  The pipeline is partially based on the
\texttt{REDCAN} pipeline for reduction of mid-IR imaging and
spectroscopy from CANARICAM on the Gran Telescopio Canarias
\citep{2013A&A...553A..35G}.  The steps began with pattern noise
cleaning, non-linearity correction, locating the spectral orders and
flat-fielding.  Sky subtractions were performed for each AB pair
closest in time, then the 2D spectra were stacked.  Spatial distortion
correction and wavelength calibrations were applied before the 1D
spectrum was extracted.  To perform telluric correction, the stellar
hydrogen lines were first removed from the telluric star spectrum.
The IRAF\footnote{The Image REduction and Analysis Facility (IRAF) is
  distributed by the National Optical Astronomy Observatories, which
  is operated by the Association of Universities of Research in
  Astronomy, Inc., under cooperative agreement with the National
  Science Foundation.} task \texttt{telluric} then interactively
adjusted the relative wavelength shift and flux scale to divide out
telluric features present in the science spectrum.  A blackbody
spectrum was then assumed for the telluric star for the flux
calibration.  Finally, the six orders were joined to form a single
continuous spectrum.

%SpeX

The SpeX spectrum was obtained in the single-prism low-resolution
(LRS) mode.  In this configuration, SpeX yields an uninterrupted
wavelength coverage from $\sim0.7$ to $2.5$ \um\ and a resolution of
$R\sim200$.  The same ABBA ``nod-along-the-slit'' technique was used
to obtain 12 frames of 150-second exposures, totaling 30 minutes of
on-target integration time.  The details in the observation and
reduction procedures can be found in \citet{2004PASP..116..362C} and
\citet{2009AJ....138..727M}.

%%%%%%%%%%
%% Phot %%
%%%%%%%%%%

\section{Photometric properties}
\label{s:phot}

%intro

The photometric properties of a SN~Ia serve as useful indicators of
possible peculiarities.  We use the preliminary Swope light curves
(without host galaxy subtractions) and pre- and post-discovery
images from P48 to establish some basic light curve parameters for
iPTF13ebh.

%explosion date

The earliest data points in the $r$-band light curve from P48 and
Swope allow for an accurate explosion time estimation.  We use a
Markov Chain Monte Carlo code to determine the time of explosion and
the associated uncertainty.  The exponent is treated as a free
parameter to allow for deviations from the $t^2$ fireball
regime \citep[e.g.,][]{2014ApJ...783L..24Z, 2014arXiv1410.1363G}.  The
time of explosion for iPTF13ebh is determined to be between the
earliest detection and the non-detection image the night before, on
2013 November 11.85 UT or
$\mathrm{JD}_{\mathrm{explosion}}=2,456,608.4\pm0.2$.  That is 0.59
day after the non-detection, 0.36 day before the first detection, 1.30
days before the discovery image, and 2.29 days before the first NIR
spectrum.  The result is robust against varying the number of light
curve points included in the calculations.  The resulting exponent is
$1.5\pm0.1$, substantially different from the fireball regime and the
average exponents determined using large samples of SN~Ia light curves
\citep[e.g.,][]{2006AJ....132.1707C, 2010ApJ...712..350H,
  2014arXiv1411.1064F}.

%testing explosion date with subtracted images

In order to test the effect of host subtraction on the time of
explosion, we subtract the 180-degree rotated image of the host from
each Swope $r$-band image.  The host, NGC~890, is morphologically
classified as a SAB galaxy \citep{1991rc3..book.....D}, and its
projected shape in the plane of the sky is assumed to have axial
symmetry (Fig.~\ref{f:image}).  The resulting subtracted $r$-band P48
and Swope light curve is again analyzed with the Markov Chain Monte
Carlo code.  The time of explosion, $2,456,608.2\pm0.2$, is fully
consistent with what we found using the subtracted P48/un-subtracted
Swope light curve.  This points to the importance of the earliest two
P48 points for the determination of time of explosion.

%explosion time from velocity evolution

\begin{figure}
\centering
\includegraphics[width=0.47\textwidth,clip=true]{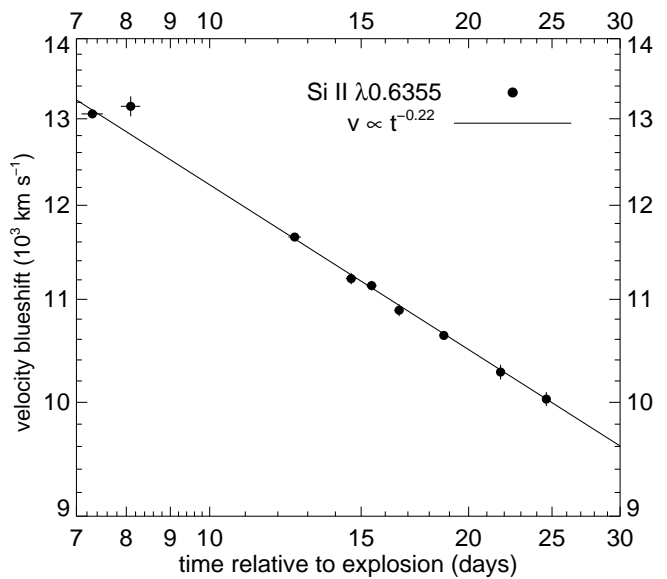}
\caption{The \siii\ \lam0.6355 \um\ velocity time evolution and the
  $v~\propto~t^{-0.22}$ power law fit of \citet{2013ApJ...769...67P}.
  The $-0.22$ exponent yields an excellent fit to the data.  The
  inferred explosion time is $19.2\pm0.2$ days before $B$ maximum.}
\label{f:vsi_texp}
\end{figure}

As a comparison to measuring the explosion time with the $r$-band
light curve, we also constrained the explosion time with a fit of the
$v~\propto~t^{-0.22}$ power law of \citet{2013ApJ...769...67P} to the
\siii\ \lam0.6355 \um\ velocity time evolution
(Fig.~\ref{f:vsi_texp}).  If this velocity-inferred explosion time is
earlier than the one measured from the light curve, it may indicate
that the supernova had a ``dark phase'' where it was initially too dim
to observe because of the delay between the time of explosion and the
time when the heating from \nirad\ first reached the outer ejecta
\citep{2013ApJ...769...67P}.  The best fit explosion time with the
$-0.22$ exponent is $19.2\pm0.2$ days before $B$ maximum.  As shown in
Fig.~\ref{f:vsi_texp}, the $-0.22$ exponent yields an excellent fit to
the data.  Treating the exponent as a free parameter yields a
consistent explosion date and the same exponent of $-0.22$.  This
inferred explosion time implies a dark phase of $\sim 4$ days and that
the \nirad\ is concentrated to the center of the ejecta.  Note that
the uncertainty in the explosion date only reflects the goodness of
the power-law fit to the data and does not take into account any
approximations used to obtain Equation 13 in
\citet{2013ApJ...769...67P}.  It is however interesting to note that
this dark phase is considerably longer than those of SNe~2009ig,
2011fe, and 2012cg considered in \citet{2014ApJ...784...85P}.

%light curve parameters

Several light-curve parameters are determined here using the Swope
optical and RetroCam NIR light curves.  The $B$-band light curve
decline rate, \dm\ \citep{1993ApJ...413L.105P} is a luminosity
indicator, but is also found to be an indicator of the strength of
some spectral features in the optical
\citep[e.g.,][]{1995ApJ...455L.147N} and NIR
\citep[e.g.,][]{2013ApJ...766...72H}.  The color-stretch parameter
\sbv, also a luminosity indicator, is found to better discriminate
among the fast-declining events than \dm\ \citep{2014ApJ...789...32B}.
Since the Swope light curves are so densely sampled and have good data
quality, we elect to interpolate the light curves using Gaussian
processes \citep{rasmussen2006}, and measure \dm, \sbv, and the time
of $B$ maximum directly.  They are listed in Table~\ref{t:basic}.
\texttt{SNooPy} fits \citep{2011AJ....141...19B} yield consistent
values for these parameters.  The time of $B$-band maximum determined
here is consistent with that found by \citet{2014ATel.5926....1W}, but
with much higher precision.  We also measured the host color excess
and the host/Milky Way extinction-corrected peak magnitudes using
\texttt{SNooPy} color-stretch fits and the intrinsic color loci
presented in \citet{2014ApJ...789...32B}.  These values are summarized
in Table~\ref{t:basic}.

\begin{table}
\caption{Summary of basic properties of iPTF13ebh}
\label{t:basic}
\centering
{\footnotesize
\begin{tabular}{rl}
\hline\hline
$\alpha$\,(J2000) & 02$^\mathrm{h}$21$^\mathrm{m}$59\fs98\\
$\delta$\,(J2000) & +33\degr16\arcmin13\farcs7\\
$\rm{JD}_{\rm{explosion,r}}$\tablefootmark{a}        & $2456608.4\pm0.2$\\
$\rm{JD}_{\rm{explosion,\siii\,v}}$\tablefootmark{b} & $2456604.2\pm0.2$\\
$\rm{JD}_{\rm{max}}$$(B)$ & $2456623.4\pm0.1$\\
\dm & $1.79\pm0.01$\\
\sbv & $0.63\pm0.02$\\
host & NGC 890\\
heliocentric redshift & $0.0133$\\
distance modulus\tablefootmark{c} & $33.63\pm0.18$\\
$E(B-V)_{\rm{host}}$ & $0.05\pm0.02$\\
$m_{u,\rm{max}}$\tablefootmark{d} & $15.29\pm0.05$\\
$m_{B,\rm{max}}$\tablefootmark{d} & $14.68\pm0.05$\\
$m_{V,\rm{max}}$\tablefootmark{d} & $14.62\pm0.04$\\
$m_{g,\rm{max}}$\tablefootmark{d} & $14.60\pm0.04$\\
$m_{r,\rm{max}}$\tablefootmark{d} & $14.64\pm0.04$\\
$m_{i,\rm{max}}$\tablefootmark{d} & $15.11\pm0.03$\\
$m_{Y,\rm{max}}$\tablefootmark{d} & $14.90\pm0.02$\\
$m_{J,\rm{max}}$\tablefootmark{d} & $14.88\pm0.02$\\
$m_{H,\rm{max}}$\tablefootmark{d} & $15.04\pm0.01$\\
$M_{u,\rm{max}}$ & $-18.34\pm0.19$\\
$M_{B,\rm{max}}$ & $-18.95\pm0.19$\\
$M_{V,\rm{max}}$ & $-19.01\pm0.18$\\
$M_{g,\rm{max}}$ & $-19.03\pm0.18$\\
$M_{r,\rm{max}}$ & $-18.99\pm0.18$\\
$M_{i,\rm{max}}$ & $-18.52\pm0.18$\\
$M_{Y,\rm{max}}$ & $-18.73\pm0.18$\\
$M_{J,\rm{max}}$ & $-18.75\pm0.18$\\
$M_{H,\rm{max}}$ & $-18.59\pm0.18$\\
\hline
\end{tabular}
}\tablefoot{\tablefoottext{a}{Derived from $r$-band light curve.}
  \tablefoottext{b}{Derived from the fit of the $v \propto t^{-0.22}$
    power law of \citet{2013ApJ...769...67P} to the \siii\ \lam0.6355
    \um\ velocity time evolution.}  \tablefoottext{c}{The distance
    modulus is derived from the host recession velocity which is
    corrected for the influence of the Virgo cluster, the Great
    Attractor, and the Shapley supercluster
    \citep{2000ApJ...529..786M}.  The error includes uncertainties
    from peculiar velocity.}  \tablefoottext{d}{The peak magnitudes
    includes reddening correction for the foreground Milky Way and the
    host extinction.}}
\end{table}

%transitional events

The \dm\ value of $1.79$ places iPTF13ebh in the category of a
fast-declining SN~Ia, but its photometric and spectroscopic properties
are not quite as extreme as subluminous or 91bg-like SNe~Ia.  Unlike
91bg-like objects, the primary maxima of the NIR $iYJH$ light curves
peak before the $B$ maximum.  The secondary maxima are present in all
$iYJH$ bands.  And its $B$ and $V$ absolute magnitudes fall within the
width-luminosity relations presented in \citet{1999AJ....118.1766P}.
It is therefore categorized as a ``transitional'' event.  In this
category, iPTF13ebh is one of the best observed and one of the most
extreme members, pushing toward the boundary between ``transitional''
and subluminous/91bg-like objects.  We explore the subject of
transitional events in more details and propose concrete definitions
for this group in Section~\ref{s:trans}.

%UV colors

\begin{figure}
\centering
\includegraphics[width=0.47\textwidth,clip=true]{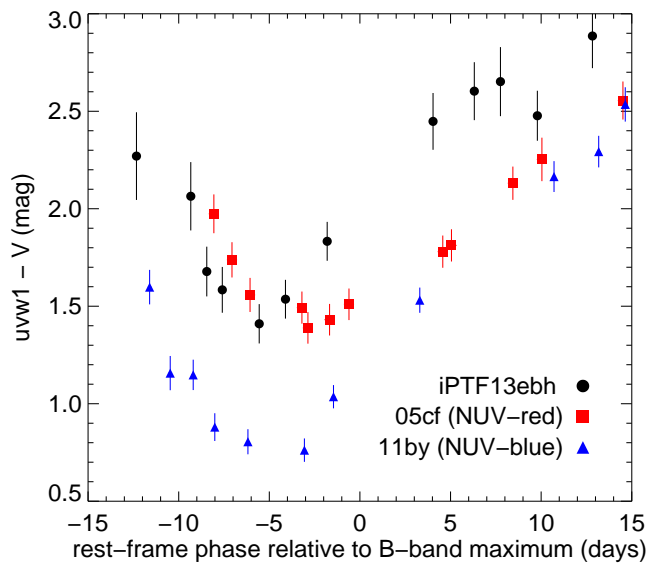}
\caption{The $uvw1-V$ color curves of iPTF13ebh, SNe~2005cf and
  2011by.  SNe~2005cf and 2011by are examples of the ``NUV-red'' and
  ``NUV-blue'' groups, respectively. The $uvw1-V$ color of iPTF13ebh
  is similar to that of the ``NUV-red'' group, but evolves faster.}
\label{f:uv_colors}
\end{figure}

In \citet{2013ApJ...779...23M}, an apparent bimodality was observed
based on UV-to-optical colors and UV spectroscopic properties.  Two
main groups were identified to be ``NUV-blue'' and ``NUV-red''.  It
was also noted that the vast majority of the SNe~Ia with optical
carbon detections (classified based on optical \cii\ lines) belong in
the ``NUV-blue'' group.  iPTF13ebh would have been classified as a
``narrow-peaked'' SN~Ia and thus excluded from the analysis of normal
SNe~Ia by \citet{2013ApJ...779...23M}.  Nonetheless, we examine the UV
color as it appears to be a strong indicator of the appearance of
carbon.  Due to the lack of UV spectra, $K$-corrections are not
applied to our UV light curves, following the same analysis as
\citet{2013ApJ...779...23M}.  The $uvw1-V$ color curve of iPTF13ebh is
quite red, staying well above $1.0$ mag before $B$ maximum
(Fig.~\ref{f:uv_colors}).  This certainly puts iPTF13ebh in the
``NUV-red'' territory.  However, note that the $uvw1-V$ color curve of
iPTF13ebh evolves faster than that of a typical ``NUV-red'' SN~Ia
(Fig.~\ref{f:uv_colors}).  This is the subject of ongoing work by
Milne et al. (in preparation).

%bias

iPTF13ebh may provide a rare example of carbon detection in a
``NUV-red'' SN~Ia.  The optical \cii\ of iPTF13ebh is weak and
disappears quickly (Section~\ref{s:c_optical}), while the NIR \ci\ is
strong early (Section~\ref{s:c_compare}).  It weakens quickly, but
persists until maximum light.  In a typical SN~Ia spectroscopic follow
up in the optical, starting more than a week past the explosion date,
the unburned carbon would have been missed.  Carbon studies based on
optical \cii\ features would then introduce a bias.  Whether this bias
results in the observed association of unburned carbon with
``NUV-blue'' SNe~Ia requires a larger sample of NIR spectra to reach a
firm conclusion.

%%%%%%%%%%%%
%% Carbon %%
%%%%%%%%%%%%

\section{NIR Carbon detections and evolution}
\label{s:carbon}

%synapps fits

The pre-maximum spectra of iPTF13ebh were analyzed using the automated
spectrum synthesis code \texttt{SYNAPPS} \citep{2011PASP..123..237T},
derived from \texttt{SYNOW} \citep{2005PASP..117..545B}.
\texttt{SYNAPPS} uses a highly parameterized, and therefore fast
spectrum synthesis technique, useful for identifying the ions that
form the observed features.  It has been employed successfully to
identify the \cii\ \lam0.6580 \um\ lines in early optical spectra
\citep{2007ApJ...654L..53T, 2011ApJ...743...27T, 2012ApJ...752L..26P}
and the \ci\ \lam1.0693 \um\ lines in NIR spectra taken near maximum
light \citep{2013ApJ...766...72H, 2015ApJ...798...39M}.

\subsection{Carbon detections in the NIR}

%NIR C I lines

\begin{figure}
\centering
\includegraphics[width=0.47\textwidth,clip=true]{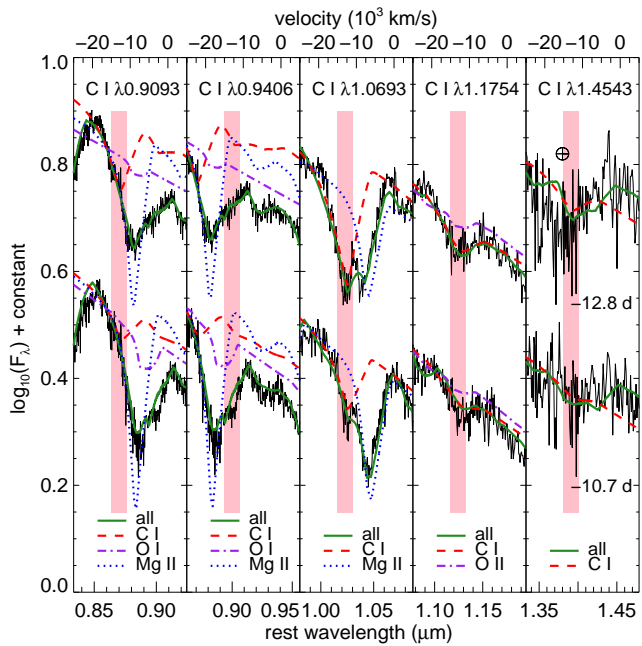}
\caption{NIR \ci\ lines of iPTF13ebh in the two earliest spectra.  The
  phase relative to the time of $B$ maximum is labeled for each
  spectrum.  The best-fit \texttt{SYNAPPS} models for both epochs
  yielded a velocity of 13,000 \kms\ for \ci, representing the base of
  the line-forming region.  The best-fit as a whole and the isolated
  contributions from the main ions responsible for the spectral
  features in each wavelength region are plotted.  A vertical band
  marks that velocity for each \ci\ line.  \ci\ \lam1.4543 \um\ for
  iPTF13ebh is located near the region of strong telluric absorption
  which is identified with an Earth symbol.}
\label{f:c_nir}
\end{figure}

In the two earliest NIR spectra of iPTF13ebh ($-12.8$ and $-10.7$ days
with respect to $B$ maximum), the \ci\ \lam1.0693 \um\ feature is
prominent.  \texttt{SYNAPPS} was employed to confirm the
identification and to analyze each NIR \ci\ line in detail.  Several
NIR \ci\ lines are identified at the velocity of $-13,000$ \kms,
representing the base of the line-forming region and are presented in
Fig.~\ref{f:c_nir}.  \ci\ \lam0.9093 \um\ is listed as the second
strongest NIR \ci\ line in Table 5 of \citet{2009AJ....138..727M} and
it shows up as weak notches in both spectra.  The \ci\ \lam\lam0.9406,
1.1754 \um\ lines appear to be present in the spectra, but some
blending makes the identification less secure.  \ci\ \lam1.4543
\um\ appears isolated from other ions; however, the proximity to the
strong telluric absorptions between the $J$ and $H$ band makes this
feature less than ideal for the search of unburned carbon.
\ci\ \lam1.0693 \um, on the other hand, forms one of the strongest
features in these early NIR spectra, even comparable to the strength
of the \mgii\ \lam1.0927 \um\ feature, which is usually the most
prominent feature in this wavelength region for SNe~Ia.

%recommendations

Echoing past works, such as \citet{2006ApJ...645.1392M}, we can
confirm here that \ci\ \lam1.0693 \um\ is the best NIR feature for
carbon detection.  All the other \ci\ lines shown here overlap with
other lines or telluric features.  For a second line to confirm the
presence of \ci, we recommend \ci\ \lam1.1754 \um.  This feature is
detected in both iPTF13ebh and SN~1999by \citep{2002ApJ...568..791H},
and tentatively identified in SN~2011fe \citep{2013ApJ...766...72H}.
It is the second strongest \ci\ line present in the NIR spectra of
SN~1999by.  The line blending effect with the \ion{O}{ii} \lam1.1667
\um\ feature is expected to be small, as neutral oxygen is the
dominant ionization state and the \ion{O}{ii} \lam1.1667 \um\ line
strength is intrinsically weak.

\subsection{Comparison to other SNe~Ia}
\label{s:c_compare}

%SN1999by

\begin{figure}
\centering
\includegraphics[width=0.47\textwidth,clip=true]{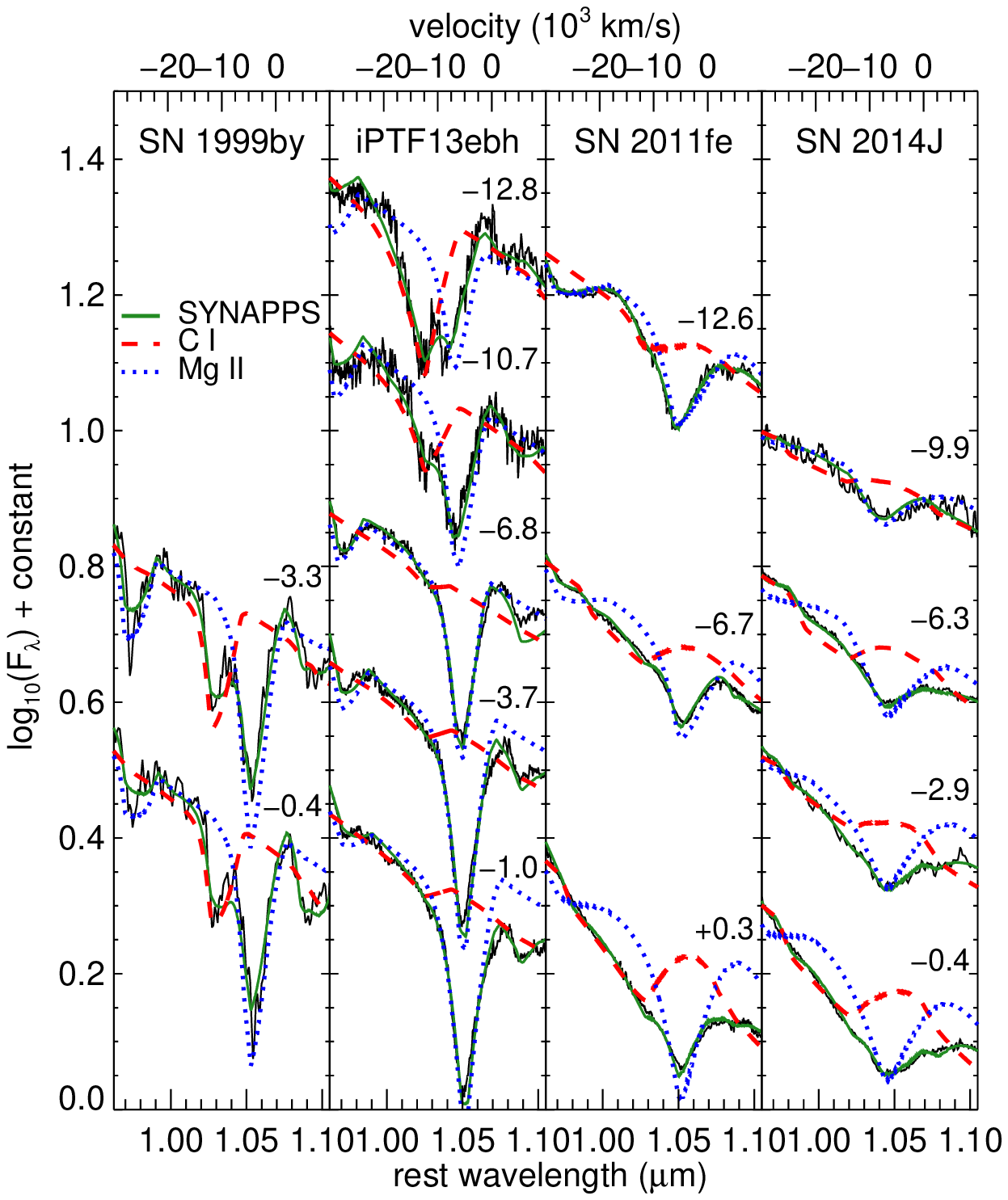}
\caption{The NIR \ci\ \lam1.0693 \um\ line for four SNe~Ia with NIR
  \ci\ detection.  Each panel presents the time evolution for each
  supernova with the phase relative to $B$ maximum labeled for each
  spectrum.  The solid green curves are the best-fit \texttt{SYNAPPS}
  models.  Dashed red and dotted blue curves show the isolated
  contributions from \ci\ \lam1.0693 \um\ and \mgii\ \lam1.0927 \um,
  respectively.  The velocity axes on the top are plotted with respect
  to the \ci\ \lam1.0693 \um\ line.}
\label{f:c_compare}
\end{figure}

So far, NIR \ci\ has been detected in only four SNe~Ia; these are
plotted in Fig.~\ref{f:c_compare} for comparison.  The first such
detection is in the 91bg-like SN~1999by.  \citet{2002ApJ...568..791H}
presented NIR spectra of SN~1999by with strong \ci\ \lam1.0693
\um\ that persisted through maximum light.  The morphology of the
feature is similar to that of iPTF13ebh, but the \ci\ \lam1.0693
\um\ feature of iPTF13ebh weakens rapidly and all but disappears by
maximum light.  The subluminous delayed detonation model of
\citet{2002ApJ...568..791H} also predicts a very strong
\ci\ \lam0.9406 \um\ line.  For the NIR spectra of SN~1999by, the
\ci\ \lam0.9406 \um\ line falls outside the observed wavelength region
so that the prediction was not tested.  The NIR spectra of iPTF13ebh
include the \ci\ \lam0.9406 \um\ feature, but as seen in
Fig.~\ref{f:c_nir}, the feature is not strong.

%SN2011fe

SN~2011fe, discovered within one day of the explosion, has been shown
to be a proto-typical SN~Ia \citep[e.g.,][]{2011Natur.480..344N,
  2012ApJ...752L..26P}.  \citet{2013ApJ...766...72H} interpreted the
flattened emission wing of \mgii\ \lam1.0092 \um\ as due to the
presence of \ci\ \lam1.0693 \um.  It was shown to increase in strength
toward maximum light (Fig.~\ref{f:c_compare}).  We note that the
profiles and the behaviors of the \ci\ features of SN~2011fe and
iPTF13ebh are drastically different.  While the \ci\ line strength
increases toward maximum light for SN~2011fe, the \ci\ line in
iPTF13ebh is strong, but decreases in strength rapidly soon after
explosion.  iPTF13ebh is the only one of the four SNe~Ia with detected
\ci\ that shows this behavior.  We note that while the first two
spectra of iPTF13ebh exhibit \ci\ profiles similar to those of
SN~1999by, subsequent spectra show only traces of \ci, manifested in
the flattened wing of \mgii\ \lam1.0092 \um, much like the \ci\ line
profile of SN~2011fe.

%SN2014J

SN~2014J developed a very similar \ci\ line profile to that of
SN~2011fe.  Indeed, as mentioned by \citet{2013ApJ...766...72H}, the
flattened wing of \mgii\ \lam1.0092 \um\ is a common characteristic of
normal-bright SNe~Ia.  \citet{2015ApJ...798...39M} showed a possible
detection of \ci\ in the \texttt{SYNAPPS} fit of the maximum-light NIR
spectrum.  Here we take advantage of the densely-cadenced observations
of SN~2014J to show the evolution of the \ci\ feature.  The
\texttt{SYNAPPS} fits in Fig.~\ref{f:c_compare} suggest a moderate
increase in strength.  Unfortunately, the late discovery of SN~2014J
\citep{2014ApJ...783L..24Z, 2014ApJ...784L..12G} did not permit early
spectroscopic follow up that is required to show definitively the
increase in \ci\ strength, as was shown in SN~2011fe
\citep{2013ApJ...766...72H}.

%SN2014J high velocity vs carbon

SN~2014J has been shown to be a normal SN~Ia, with significant
reddening associated with a large amount of dust
\citep{2014ApJ...788L..21A, 2014MNRAS.443.2887F, 2014arXiv1408.2381B},
and relatively high velocity for \siii\ \lam0.6355
\um\ \citep{2015ApJ...798...39M} that places the supernova near the
boundaries between the normal and high-velocity sub-classes
\citep[e.g.,][]{2005ApJ...623.1011B, 2009ApJ...699L.139W}. In large
samples of optical spectra, \citet{2012ApJ...745...74F} and
\citet{2012MNRAS.425.1917S} both observed that SNe~Ia with carbon tend
to have lower \siii\ velocities, while the objects without carbon span
the entire range of \siii\ velocities.  This phenomenon may be a
result of the difficulty in detecting \cii\ \lam0.6580 \um\ at high
velocities where it overlaps with the \siii\ \lam0.6355 \um\ line.
The detection of NIR \ci\ \lam1.0693 \um\ in the moderately-high
velocity SN~2014J shows that searching for signatures of unburned
carbon in the NIR may remedy this observational bias.  We explore the
differences in NIR \ci\ and optical \cii\ lines further in the
following subsection.

%C I velocity

\begin{figure}
\centering
\includegraphics[width=0.47\textwidth,clip=true]{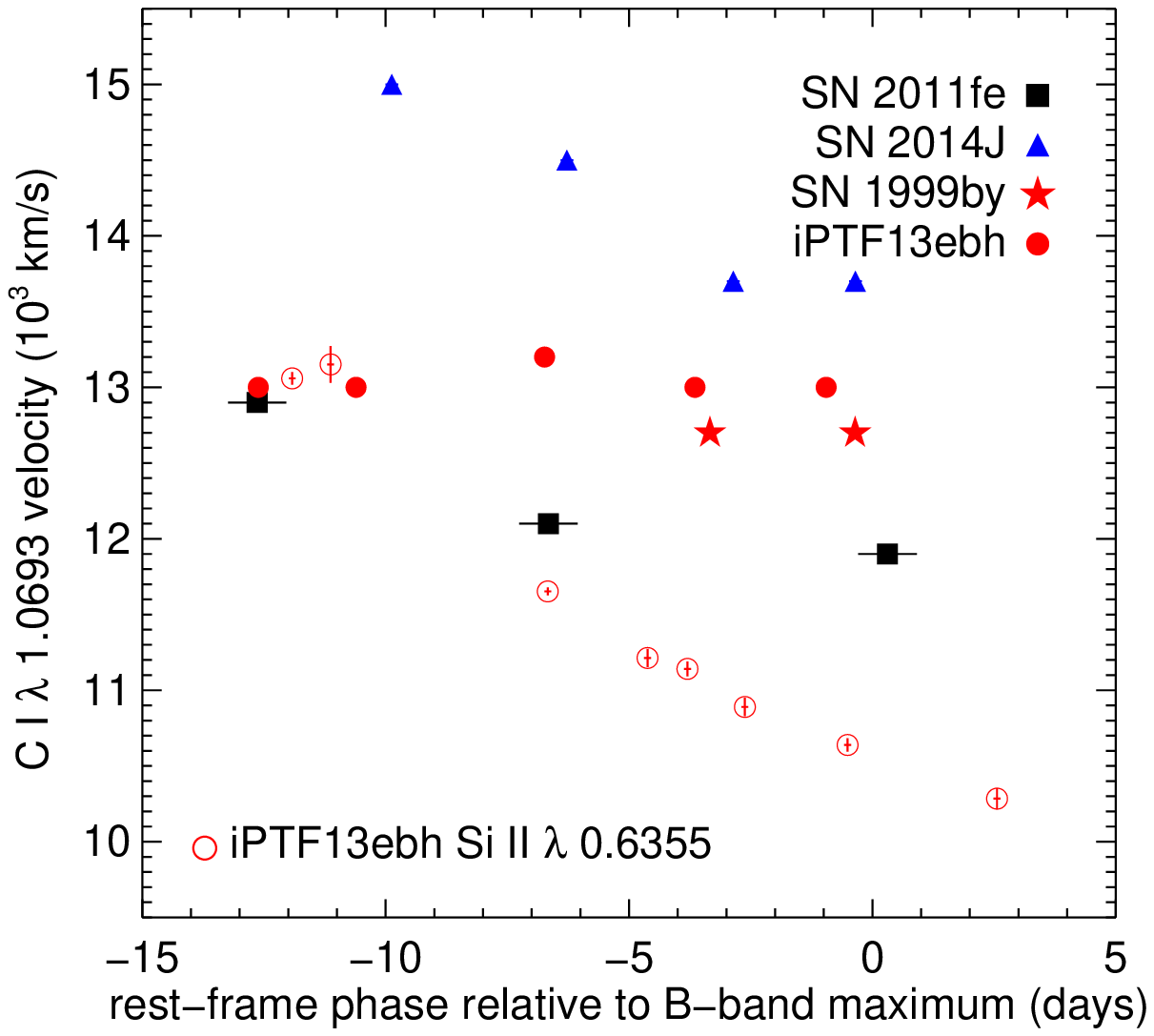}
\caption{The time evolution of the \ci\ \lam1.0693 \um\ velocities for
  the four supernovae with \ci\ detections.  Various symbols represent
  different objects as noted.  Blue, black, and red symbols represent
  SNe~Ia with optical light-curve decline rates in the range \dm<1.0,
  1.0<\dm<1.6, \dm>1.6, respectively.  The velocities are taken from
  the \ci\ velocity representing the base of the line-forming region
  in the best-fit \texttt{SYNAPPS} models.  The \siii\ \lam0.6355
  \um\ velocity of iPTF13ebh is also plotted for comparison.}
\label{f:c_v}
\end{figure}

In Fig.~\ref{f:c_v}, we compare the \ci\ velocity evolution for the
four SNe~Ia with \ci\ detections.  As the location of the \ci\ line
minimum in most cases cannot be directly measured
(Fig.~\ref{f:c_compare}), we plot the \ci\ velocities, representing
the base of the line-forming region from the best-fit \texttt{SYNAPPS}
models.  They therefore have no associated uncertainties.  For both
iPTF13ebh and SN~1999by, the \ci\ velocity evolution is flat
(although, SN~1999by does not have early-phase data to rule out an
early decline).  The line-forming region probes the inner carbon-rich
layers.  For SNe~2011fe and 2014J, the \ci\ velocities continue to
decline until only a few days before maximum.
\citet{2011ApJ...732...30P} presented the \cii-to-\siii\ velocity
ratio measured from a sample of optical spectra and showed flat time
evolution and a ratio of $\sim 1.1$ for the majority of the SNe~Ia
studied.  iPTF13ebh certainly falls outside of this norm with varying
\ci-to-\siii\ velocity ratio increasing from $\sim1$ to $\sim1.3$.

\subsection{Comparison of NIR \ci\ and optical \cii}
\label{s:c_optical}

%intro

The search for unburned material in SNe~Ia has largely been focused at
optical wavelengths and principally on the \cii\ \lam0.6580
\um\ feature, which disappears quickly a few days past explosion.  The
recent detections of the NIR \ci\ \lam1.0693 \um\ line, still present
near maximum light, suggest that going to the NIR is a less biased way
of investigating unburned material.  In this section, we compare the
NIR \ci\ and optical \cii\ features for the four SNe~Ia with reported
NIR \ci\ detection.

%comparison for iPTF13ebh

\begin{figure*}
\centering
\includegraphics[width=0.95\textwidth,clip=true]{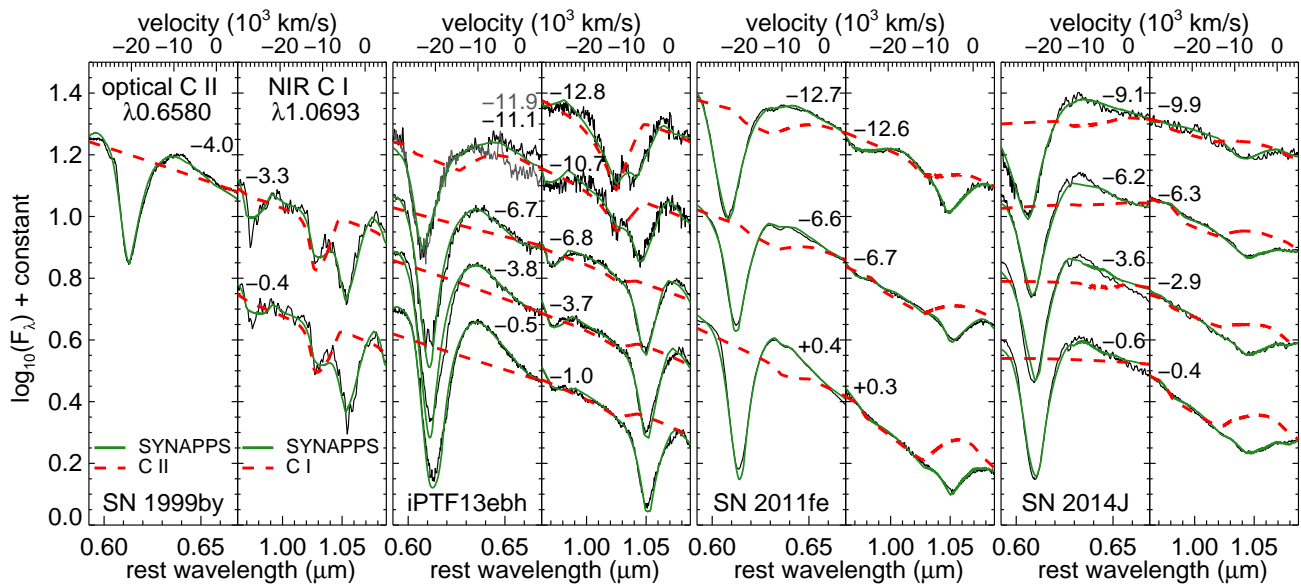}
\caption{Comparison between the optical \cii\ \lam0.6580 \um\ and the
  NIR \ci\ \lam1.0693 \um\ lines of the four SNe~Ia with
  \ci\ detections.  The optical/NIR pair is selected such that the
  spectra are close in phase for each SN~Ia.  The phases relative to
  $B$ maximum are labeled.  The velocity axes are plotted with respect
  to each carbon line.  Since the two earliest optical spectra of
  iPTF13ebh are taken less than one day apart and have identical
  \siii/\cii\ line profiles, they are plotted together.  The
  \texttt{SYNAPPS} fit was done on the day $-11.1$ spectrum.  Except
  at the very early epoch of SN~2011fe, the NIR \ci\ line is always
  stronger than the optical \cii\ line.}
\label{f:c_opt_time}
\end{figure*}

In Fig.~\ref{f:c_opt_time}, we compare the NIR \ci\ and optical
\cii\ features of iPTF13ebh as inferred by the \texttt{SYNAPPS}
models.  As mentioned in the previous subsection, the NIR \ci\ of
iPTF13ebh is the strongest ever observed, but weakens rapidly in
dramatic contrast to the behavior of the NIR \ci\ lines of SNe~2011fe
and 2014J.  The first two NIR spectra, taken at $-12.8$ and $-10.7$
days relative to $B$ maximum, both show a strong \ci\ \lam1.0693
\um\ line, and there is evidence for a weak \cii\ \lam0.6580 \um\ line
in the first two optical spectra taken at $-11.9$ and $-11.1$ days
relative to $B$ maximum.  In the classification scheme of
\citet{2012ApJ...745...74F}, the \cii\ \lam0.6580 \um\ feature would
be classified as ``F'' for flat \siii\ \lam0.6355 \um\ emission, and
not as ``A'' for a definitive carbon absorption.  The optical
\cii\ line also weakens rapidly.  Both the weak optical \cii\ and the
strong NIR \ci\ features appear at approximately the same epoch.  It
is possible that the optical \cii\ feature was stronger at an even
earlier epoch, then later yielded to the growing \ci\ due to
recombination.

%comparison for other SNe

We now consider the optical \cii\ lines for other SNe~Ia with NIR
\ci\ detections.  The optical spectra of SN~1999by from
\citet{2004ApJ...613.1120G} and \citet{2008AJ....135.1598M} showed no
strong \cii\ features from $-5$ days through to maximum light, while
the NIR spectra of \citet{2002ApJ...568..791H} showed strong
detections of several \ci\ features during the same phases.  The
optical \cii\ \lam0.6580 \um\ feature was detected in SN~2011fe
\citep{2012ApJ...752L..26P}, and is weaker than the NIR \ci,
especially at maximum light \citep{2013ApJ...766...72H}.  The NIR
\ci\ is weak early on but gradually increases in strength up until
maximum light.  While \ci\ was detected in several pre-maximum NIR
spectra of SN~2014J \citep{2015ApJ...798...39M}, \cii\ was not
detected in the optical spectra taken during the same phases
\citep{2014ApJ...784L..12G}.  These optical \cii\ and NIR
\ci\ comparisons are summarized in Fig.~\ref{f:c_opt_time}, along with
the \texttt{SYNAPPS} fits and the illustration of the \ci\ and
\cii\ contributions.

%conclusion

While the line profile, strength and time evolution of the NIR
\ci\ lines are drastically different in iPTF13ebh, SN~1999by, and
normal SNe~Ia like SNe~2011fe and 2014J, the common theme is: except
at the very early epochs, the strength of the NIR \ci\ lines is {\em
  always} observed to be much stronger than that of their optical
\cii\ counterparts (Fig.~\ref{f:c_opt_time}).  The laboratory line
strength of \cii\ \lam0.6580 \um\ is a few orders of magnitude
stronger than that of \ci\ \lam1.0693 \um\ at a reasonable range of
temperature.  The observations of stronger NIR \ci\ lines indicate
that SN~Ia, regardless of their luminosity, produce conditions in the
outer layers that are favorable to a much higher abundance of
\ci\ than \cii.  This appears to be in contradiction with the
prediction by \citet{2008ApJ...677..448T}, that the dominant carbon
ionization state in the outer parts of the ejecta is \cii.

%caveats

Four \ci\ detections is obviously too small a sample to reach a firm
conclusion.  However, we note here that the small number of
\ci\ detections is not due to the lack of or weak \ci\ in SNe~Ia, but
the the small sample size of pre-maximum NIR spectra.  The most recent
analyses of SN~Ia NIR spectroscopy have all revealed \ci\ \lam1.0693
\um\ to be present \citep{2013ApJ...766...72H, 2015ApJ...798...39M,
  2015A&A...573A...2S}

%%%%%%%%%%%%%
%% Spectra %%
%%%%%%%%%%%%%

\section{Other Spectroscopic Properties}
\label{s:spec}

%intro

Besides probing unburned material in the ejecta, early time-series NIR
spectra provide measures of several physical parameters in SN~Ia
explosions \citep{1998ApJ...496..908W}.  \citet{2013ApJ...766...72H}
introduced quantitative measurements for the $H$-band break and
\mgii\ velocity.  We explore these quantities and other spectroscopic
properties of iPTF13ebh in this section.

\subsection{Optical Spectroscopic Properties}
\label{s:spec:opt}

%pW

Pseudo-equivalent widths ($pEW$) have been widely adopted as a tool
for quantifying supernova spectral features.  We measure the $pEW$ of
several optical spectral features of iPTF13ebh near maximum light, in
the same fashion as \citet{2013ApJ...773...53F}.  In particular, the
$pEW$ measurements of the \siii\ \lam\lam 0.5972, 0.6355 \um\ lines
are $48.9\pm0.6$ \AA\ and $125.2\pm0.5$ \AA, respectively.  First
introduced by \citet{1995ApJ...455L.147N}, the $R(\siii)$ ratio is
defined as the ratio of the depth of the \siii\ \lam\lam 0.5972,
0.6355 \um\ lines, and was shown to correlate with the peak
luminosities of SNe~Ia.  A similar correlation had been shown between
the $pEW$ ratio of the same features and light-curve decline rate
\citep{2012AJ....143..126B, 2013ApJ...773...53F}.  The \siii\ $pEW$
ratio of iPTF13ebh fits well with other fast-declining SNe~Ia in this
correlation.

%Branch classification

\begin{figure}
\centering
\includegraphics[width=0.47\textwidth,clip=true]{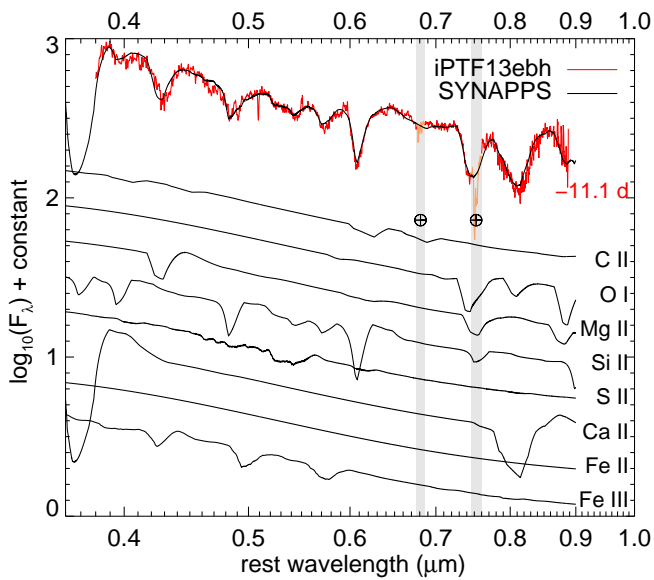}
\caption{\texttt{SYNAPPS} fit of the early optical spectrum of
  iPTF13ebh, taken at $11.1$ days before $B$ maximum.  Isolated
  contributions from ions included in the fit are also shown.}
\label{f:opt_synapps}
\end{figure}

Using the definition given by \citet{2013ApJ...773...53F}, iPTF13ebh
is placed firmly in the ``cool'' category of
\citet{2006PASP..118..560B}, for $pEW(\siii\,0.5972) > 30$ \AA.  The
maximum light absorption complex near 0.43 \um\ is attributed to
\mgii\ and \feiii\ in normal SNe~Ia.  This feature is dominated by
strong \tiii\ absorptions in the spectra of 91bg-like events.
\citet{2013ApJ...773...53F} named the $pEW$ of this feature $pEW3$ and
defined ``extreme cool'' (eCL) events to have $pEW3>220$ \AA.  This
group of SNe~Ia is largely comprised of 91bg-like events, where heavy
\tiii\ absorptions increase $pEW3$ drastically.  For iPTF13ebh, $pEW3$
is measured to be $104\pm1$ \AA, confirming the weakness of \tiii.
\texttt{SYNAPPS} fits are also performed on the pre-maximum optical
spectra of iPTF13ebh.  The complex near 0.43 \um\ described above is
attributed to \mgii, \feiii, and to a lesser extent,
\siiii\ (Fig.~\ref{f:opt_synapps}).  No \tiii\ contribution is
required to account for the observed profiles.

%Detached Ca II

High-velocity \caii\ triplet absorption features have been a topic of
interest.  Interpreted as a density enhancement in the outer region,
the presence of high-velocity \caii\ could be a signature of
circumstellar material bound in the progenitor system which was swept
up during the explosion \citep{2004ApJ...607..391G}.
\citet{2005ApJ...623L..37M} observed that almost all SNe~Ia with early
spectra show high-velocity \caii.  The question is: for the few SNe~Ia
that do not show high-velocity \caii, would they show high-velocity
\caii\ if an early spectrum were available?  iPTF13ebh is discovered
exceptionally young with the first optical spectrum obtained 3.2 days
past explosion.  (The \caii\ triplet is present in the NIR spectra,
but the detached high-velocity component, if present, would be outside
of the wavelength coverage.)  \texttt{SYNAPPS} fits for all
pre-maximum optical spectra show that only a photospheric component is
required to yield excellent fits to the observed \caii\ triplet near
0.8 \um. (See Fig.~\ref{f:opt_synapps} for example.)  For iPTF13ebh,
there is no evidence of a detached high-velocity \caii\ feature as
early as 3.2 days past explosion.  This is in agreement with previous
results that low-luminosity SNe~Ia are less likely to show strong
detached high-velocity \caii\ features at early times
\citep{2009PhDT.......228H, 2014MNRAS.437..338C, 2014MNRAS.444.3258M}.

\subsection{Delayed Detonation Model}

%intro

\citet{2002ApJ...568..791H} constructed a set of delayed detonation
\citep[e.g.,][]{1991A&A...245..114K} models that span the range of
normal-bright to subluminous SNe~Ia.  For iPTF13ebh, we select the
model with a deflagration-to-detonation transition density of
$\rho_{tr}=16\times10^6$ g cm$^{-3}$, which synthesized 0.268
$\mathrm{M}_{\odot}$ of \nirad.  The choice is based on the peak
absolute brightness and decline rate of iPTF13ebh, with no further
fine tuning to match the model to the observed spectra.  The model
yields a peak absolute magnitude in $B$ of $-18.22$ mag, a decline
rate in $B$ of \dm$=1.82$, and the rise time between the time of
explosion and $B$ maximum of 15.4 days, comparable to the observed
values for iPTF13ebh (Table~\ref{t:basic}).

%atomic data

For the calculations of the explosions, light curves, and spectra, we
use the radiation transport code HYDRA, which includes hydrodynamic
solvers, Eddington tensor and Monte Carlo methods for low-energy
photon, gamma-ray and positron transport, and time-dependent, nuclear
and atomic networks \citep{1990A&A...229..191H, 1995ApJ...444..831H,
  2002astro.ph..7103H, 2009AIPC.1171..161H, 2014ApJ...795...84P}.  The
synthetic light curves and spectra were recomputed using updated
atomic data from \citet{2010MNRAS.401.2081D} and
\citet{2014ApJ...792..120F} as described in
\citet{2014arXiv1410.6759D} and \citet{2015ApJ...798...93T}.

%early phases

\begin{figure*}
\centering
\includegraphics[width=0.90\textwidth,clip=true]{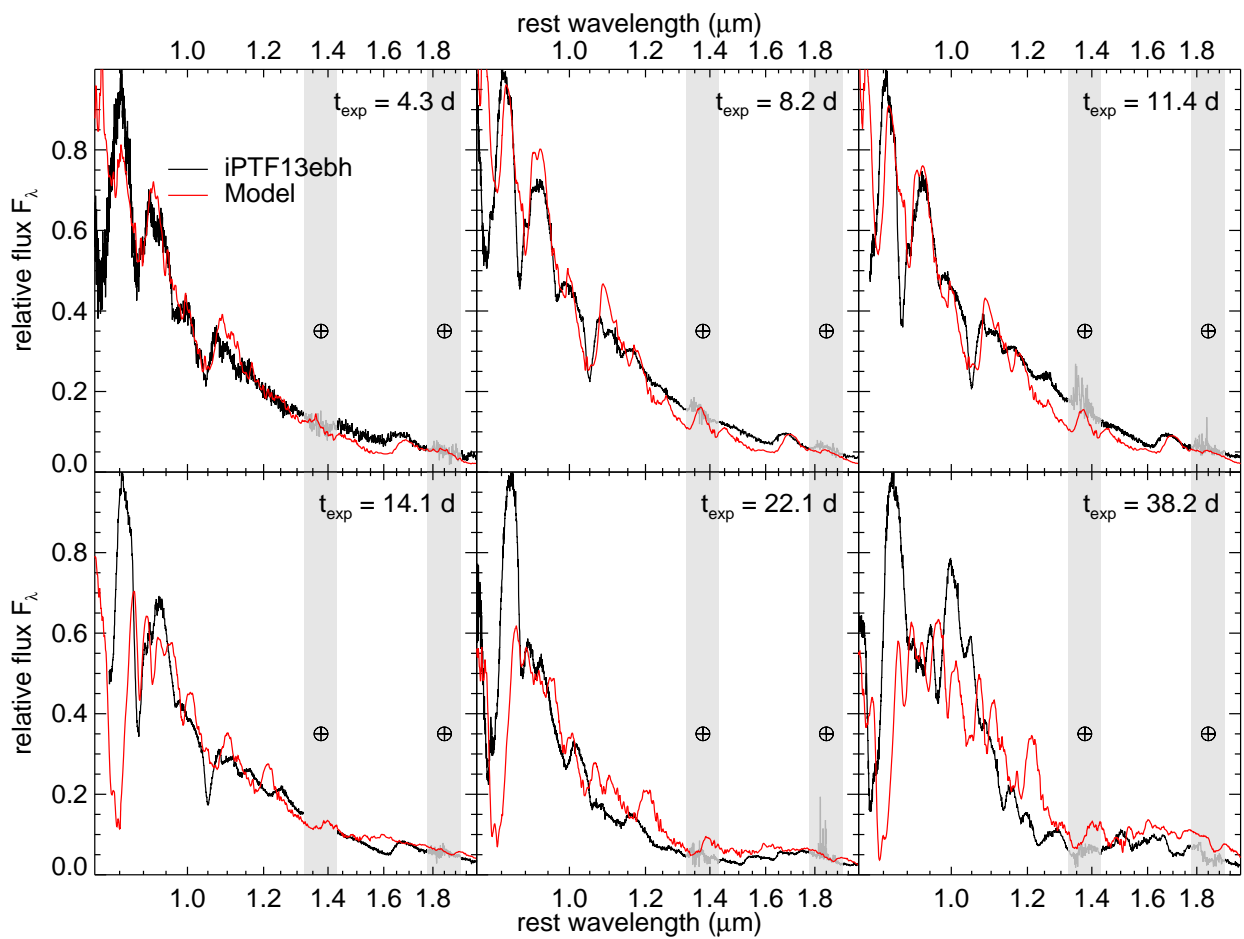}
\caption{Comparison between the NIR spectra of iPTF13ebh and the
  $\rho_{tr}=16\times10^6$ g cm$^{-3}$ model spectra.  The phases of
  the observed spectra are noted on the plot.  These are relative to
  the explosion date, as inferred from the early-time $r$-band light
  curve.  The model spectra are computed at these epochs.}
\label{f:dd_spec}
\end{figure*}

In Fig.~\ref{f:dd_spec}, the iPTF13ebh and the
$\rho_{tr}=16\times10^6$ g cm$^{-3}$ model spectra are compared
between 4 and 38 days past explosion.  The line identifications are
similar to those given in \citet{1998ApJ...496..908W} and
\citet{2002ApJ...568..791H}.  At early epochs, the spectra are
dominated by blends of \ci, \oi, \mgii, \siii, \caii, and
singly-ionized elements in the iron group.  The strong feature at
$\sim0.9$ \um\ is identified as \mgii\ \lam\lam0.9218, 0.9244 \um,
\oi\ \lam0.9266 \um, and \ci\ \lam\lam0.9406, 0.9658 \um.  And the
feature at $\sim1.05$ \um\ is identified as \ci\ \lam1.0693 \um\ and
\mgii\ \lam\lam1.0914, 1.0951 \um\ (named \mgii\ \lam1.0927 \um\ in
other sections).  The feature at $\sim1.35$ \um\ is identified as
\siii\ \lam1.3650 \um.  Between 8 and 11 days after the explosion (top
panels of Fig.~\ref{f:dd_spec}), the feature at 1.05 \um\ becomes too
broad with the emission wing too strong in the model compared to the
observations.  This feature is produced by \mgii, with contributions
from \ci\ on the blue side and the suppression of the emission
component due to iron-group elements on the red side.  Increasing the
\nirad\ abundance in the outer ejecta would have two desired effects:
higher ionization which would lead to decreased contribution from \ci,
and a stronger suppression on the emission wing from the iron-group
elements.  These may be achieved through a brighter model, stronger
outward mixing of \nirad, or higher progenitor metallicity.  Note that
the ionization of carbon is sensitive to changes in transition density
only for models in this range of brightness.

%late phases

In general, the influence of iron-peak elements increases with time,
as the spectrum-formation region probes deeper ejecta layers.
However, the 1.05 \um\ region remains relatively clear of iron-peak
lines before maximum, demonstrating again that the \ci\ and
\mgii\ lines here are ideal for studying unburned material and the
boundary between carbon and oxygen burning.  At later time, around and
after $B$ maximum (bottom panels of Fig.~\ref{f:dd_spec}), the model
features have much lower velocities compared to iPTF13ebh, indicating
that the model has a line-forming region that recedes too fast.  This
suggests a need for an increase in the model \nirad\ production.
Alternatively, the outward mixing of \nirad\ may increase the opacity
of layers of partial burning.  Note that using the the explosion date
inferred from a power-law fit to the \siii\ \lam0.6355 \um\ velocity
measurements improves the situation, but the evolution of the NIR
spectral features is still much too fast in the model spectra.

%carbon

At early time, less than a week past explosion, the model produces
\ci\ lines that are slightly weaker than ones observed in iPTF13ebh.
The ionization balance of \ci\ and \cii\ depends sensitively on the
recombination rate and heating.  \ci\ could become stronger by
enhanced density as a result of interaction with circumstellar
material in the progenitor system \citep{2004ApJ...607..391G} which
increases the recombination rate, or by cooling due to early CO
formation \citep{1995ApJ...444..831H}.  The strength of the model
\ci\ lines can also be increased by lowering the \nirad\ mass.
However, the fast-evolving model spectra mentioned above suggests that
lowering the model \nirad\ would cause further discrepancy between the
model and observed spectra.

\subsection{$H$-band Iron-peak Feature}
\label{s:hbreak}

%intro

The $H$-band iron-peak feature, a complex formed by \feii/\coii/\niii,
is the most prominent spectral feature for a normal SN~Ia in the NIR.
First noted by \citet{1973ApJ...180L..97K}, it is fortuitously located
in the $H$ band, in between the two strong telluric regions in the
NIR, such that it is relatively well documented for nearby objects.
As the interest in SN~Ia cosmology in the NIR grows, understanding
this spectral feature, which shifts out of the $H$-band for distant
objects, becomes especially important for $K$-correction calculations
\citep{2013ApJ...766...72H, 2014PASP..126..324B}.  The $H$-band
iron-peak feature also holds the promise to provide insight into the
explosion, as the strongly-variable opacity provides views of very
different depths at the same epoch \citep{1998ApJ...496..908W}.

%H-band ratio vs phase

\begin{figure}
\centering
\includegraphics[width=0.47\textwidth,clip=true]{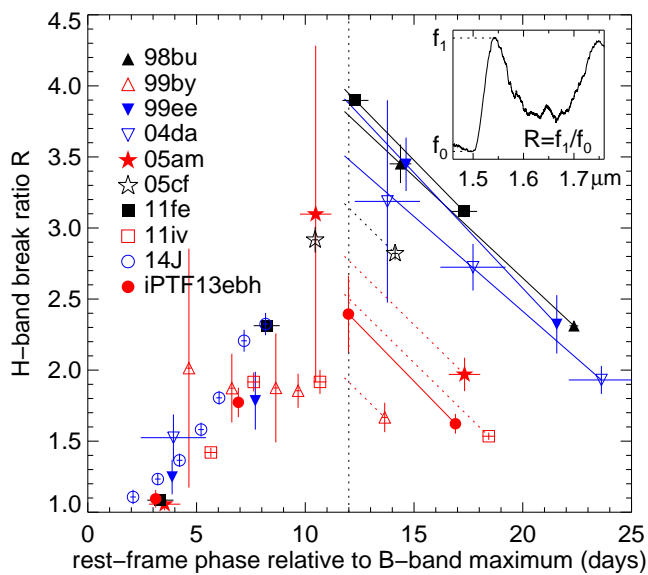}
\caption{Time evolution of the $H$-band break ratio.  Different
  symbols represent measurements of different supernovae, as noted in
  the plot.  Blue, black, and red symbols represent SNe~Ia with
  optical light-curve decline rates in the range \dm<1.0, 1.0<\dm<1.6,
  \dm>1.6, respectively.  The inset illustrates the definition of the
  $H$-band break ratio with the spectrum of SN~2011fe at 12 days past
  maximum.  The linear fits to the post-peak decline are plotted in
  solid lines when more than two data points are available and in
  dashed lines when only a single data point is available and the
  error-weighted mean decline rate is assumed.  There is a large range
  of peak $H$-band break ratios, but the post-peak declines are
  remarkably uniform.}
\label{f:hbreak_t}
\end{figure}

\citet{2013ApJ...766...72H} defined a quantitative measure for the
size of this feature.  They measured the peak-to-trough flux ratios
across the ``$H$-band break'' near 1.5 \um.  Here, we reproduce
Fig.~10 of \citet{2013ApJ...766...72H} in Fig.~\ref{f:hbreak_t}, and
now include measurements for iPTF13ebh and of the recently published
NIR spectra of SN~2014J \citep{2015ApJ...798...39M}.  The flux ratios
are measured consistently across the entire sample.  As noted in
\citet{2013ApJ...766...72H}, if we exclude the data points from the
peculiar SN~1999by, the rise and decline of the $H$-band break ratio
is remarkably uniform.  The high-cadence observations of SN~2014J show
the details of the uniform monotonic rise, although the starting point
appears to have occurred one day earlier than for the other SNe~Ia.

%iPTF13ebh

The budding of the $H$-band feature for iPTF13ebh occurs around 3 days
past maximum light, which coincides with several other SNe~Ia,
including SN~2011fe.  There is a large range of peak ratios; however,
the post-peak decline rate of the $H$-band break ratio appears to be
quite uniform.  Assuming linear declines, the decline rates of the
$H$-band break ratio for SN~2011fe and iPTF13ebh are identical within
the uncertainties, $-0.16\pm0.01$ day$^{-1}$ and $-0.16\pm0.06$
day$^{-1}$, respectively, even though the two supernovae have
drastically different luminosities.  The error-weighted mean decline
rate determined by \citet{2013ApJ...766...72H} is $-0.15\pm0.04$
day$^{-1}$.  This value appears to be robust among SNe~Ia with a large
range of luminosities.  SN~2011iv, another fast-declining transitional
object (\dm$=1.76\pm0.02$), shows some peculiarity in the time
evolution of the flux ratio.  The flux ratio appears to peak earlier
than the normal 12 days past maximum inferred from available data,
however the cadence is not dense enough to definitively locate the
peak of the $H$-band break ratio.

%correlation with dm15

\begin{figure}
\centering
\includegraphics[width=0.47\textwidth,clip=true]{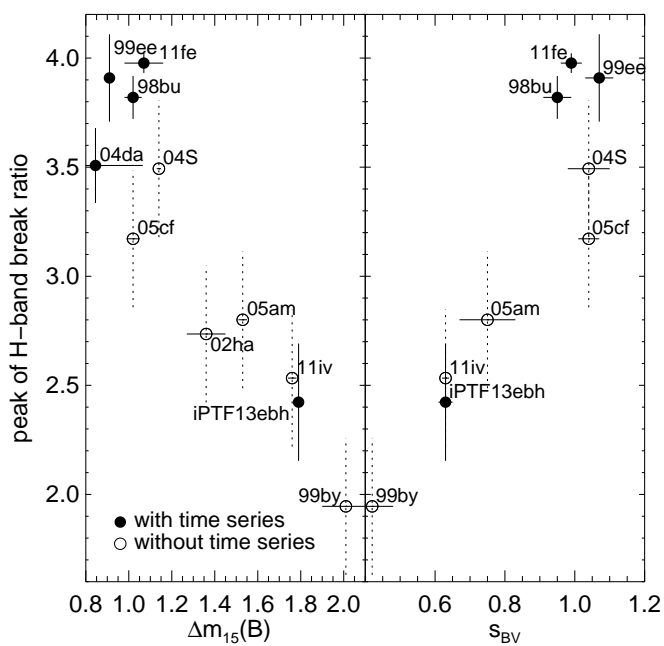}
\caption{The peak of $H$-band break ratio versus the optical
  light-curve decline rate \dm\ and color stretch \sbv. SNe~Ia with
  two or more observations in the post-peak decline are plotted with
  filled circles and solid error bars, while the rest are plotted with
  open circles and dotted error bars.  The correlations for both
  parameters are strong.}
\label{f:hbreak_dm15}
\end{figure}

\citet{2013ApJ...766...72H} noted the strong correlation between the
peak $H$-band break ratio and \dm.  In Fig.~\ref{f:hbreak_dm15}, we
reproduce Fig.~11 of \citet{2013ApJ...766...72H} using the same
technique, with the addition of iPTF13ebh and a few other SNe~Ia.  We
also plot the $H$-band ratio against the color-stretch parameter
\sbv\ (see \citet{2014ApJ...789...32B} and Section~\ref{s:trans}).
From Fig.~\ref{f:hbreak_t} and following the same procedure as
\citet{2013ApJ...766...72H}, we assume that the time evolution of the
$H$-band ratio reaches its peak at 12 days past $B$ maximum and
declines linearly thereafter.  For objects with two or more post-peak
measurements, the directly-measured decline rate is used to
extrapolate to the peak.  For SNe with one post-peak measurement, we
assume the error-weighted mean decline rate of $-0.15\pm0.04$
day$^{-1}$.  The correlation in Fig.~\ref{f:hbreak_dm15} remains quite
strong for both light-curve parameters, \dm\ and \sbv, with the
addition of a few more objects.

%why the correlation

The correlation between the $H$-band feature and light-curve decline
rate (and luminosity) was first pointed out by
\citet{2009PhDT.......228H}.  Using principal component analysis (PCA)
on eight published NIR spectra, the PCA model showed that the SN
luminosity is not only correlated with the emission strength of the
$H$-band feature, but also with its width.  Low-luminosity and
fast-declining SNe~Ia have weaker $H$-band features and have the
iron-peak material responsible for the feature confined at lower
velocities.  Within the Chandrasekhar mass delayed detonation scenario
\citep{1991A&A...245..114K, 1992ApJ...393L..55Y}, these correlations
are expected, as the strength and width of the $H$-band
features are indicators of the amount of
\nirad\ \citep{1998ApJ...496..908W, 2002ApJ...568..791H}.  On the
other hand, for explosion scenarios which produce a spread of
explosion masses such as dynamical mergers \citep{1984ApJS...54..335I,
  1984ApJ...277..355W}, or in the case of strong mixing, the
correlations are expected to be weak.

%profile comparison

\begin{figure}
\centering
\includegraphics[width=0.47\textwidth,clip=true]{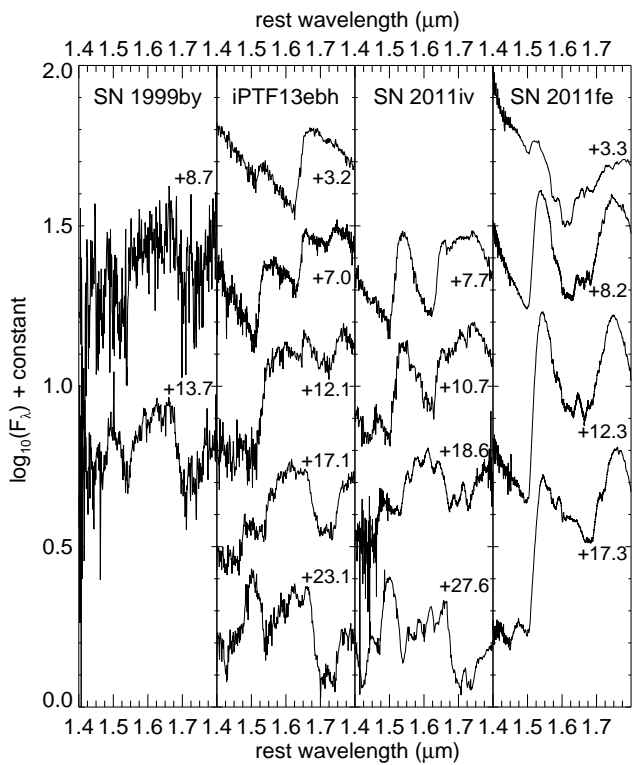}
\caption{Comparison of the $H$-band iron-peak feature between four
  SNe~Ia.  The phase with respect to $B$-band maximum is labeled for
  each spectrum.}
\label{f:hbreak_compare}
\end{figure}

The $H$-band break ratio measurements greatly simplify the analysis of
the complex profile shape and time evolution of the $H$-band feature.
In Fig.~\ref{f:hbreak_compare}, we plot the time series of the
$H$-band feature of four SNe~Ia for comparison.  SN~2011fe shows the
$H$-band break forming at approximately 3 days past maximum and
rapidly growing in strength.  The characteristic double-peaked
profile, formed by \feii/\coii/\niii, is typical of a normal SN~Ia.
iPTF13ebh and SN~2011iv are both low-luminosity, fast-declining
objects in the ``transitional'' category, yet the profile evolution of
their $H$-band features is subtly different.  At $1-2$ weeks past
maximum, the local minimum near 1.65 \um\ is more pronounced in
SN~2011iv than in iPTF13ebh.  The lack of a local minimum near 1.65
\um\ is a characteristic of the subluminous model presented by
\citet{2002ApJ...568..791H}.  By $\sim$18 days past maximum, the
$H$-band breaks for both iPTF13ebh and SN~2011iv have shifted redward
in wavelength, due to the decreasing size of the emission region at
late times.  The latest spectra shown in Fig.~\ref{f:hbreak_compare}
for iPTF13ebh, SNe~1999by and 2011iv share very similar profile
shapes, with the appearance of a strong \feii/\coii\ feature near 1.5
\um, which is usually weak in normal-bright SNe~Ia.  SN~1999by evolved
to this profile shape approximately two weeks earlier than iPTF13ebh
and SN~2011iv.

%future studies

Increasing central density causes a larger fraction of material to
undergo electron capture during deflagration and produces increasing
amounts stable \nista\ \citep{2006NuPhA.777..579H}.  Using the time
evolution of the detailed profile shape of the $H$-band complex, it is
possible to isolate features produced by stable and unstable isotopes
of iron-group elements and extract their mass ratio.  This will be a
subject of future studies with a larger data set.

\subsection{Magnesium velocity}

%intro

Magnesium is a product of explosive carbon burning, and not oxygen
burning.  This makes magnesium lines very sensitive to the
location of the inner edge of carbon burning in velocity space
\citep{1998ApJ...496..908W, 2001RMxAC..10..190M, 2002ApJ...568..791H,
  2009AJ....138..727M}.  The strong and isolated NIR \mgii\ \lam1.0927
\um\ line is ideal for such a measurement.  With early NIR
spectroscopy of SN~2011fe, \citet{2013ApJ...766...72H} showed that the
\mgii\ velocity declines rapidly and flattens between $-10$ to $10$
days past maximum.  The flattened \mgii\ velocity appears to be
ubiquitous among normal SNe~Ia.  This means that it is possible to use
a single ``snapshot'' NIR spectrum between $-10$ to $10$ days past
maximum to derive a characteristic minimum velocity for the given
object.

%MgII vs phase

\begin{figure}
\centering
\includegraphics[width=0.47\textwidth,clip=true]{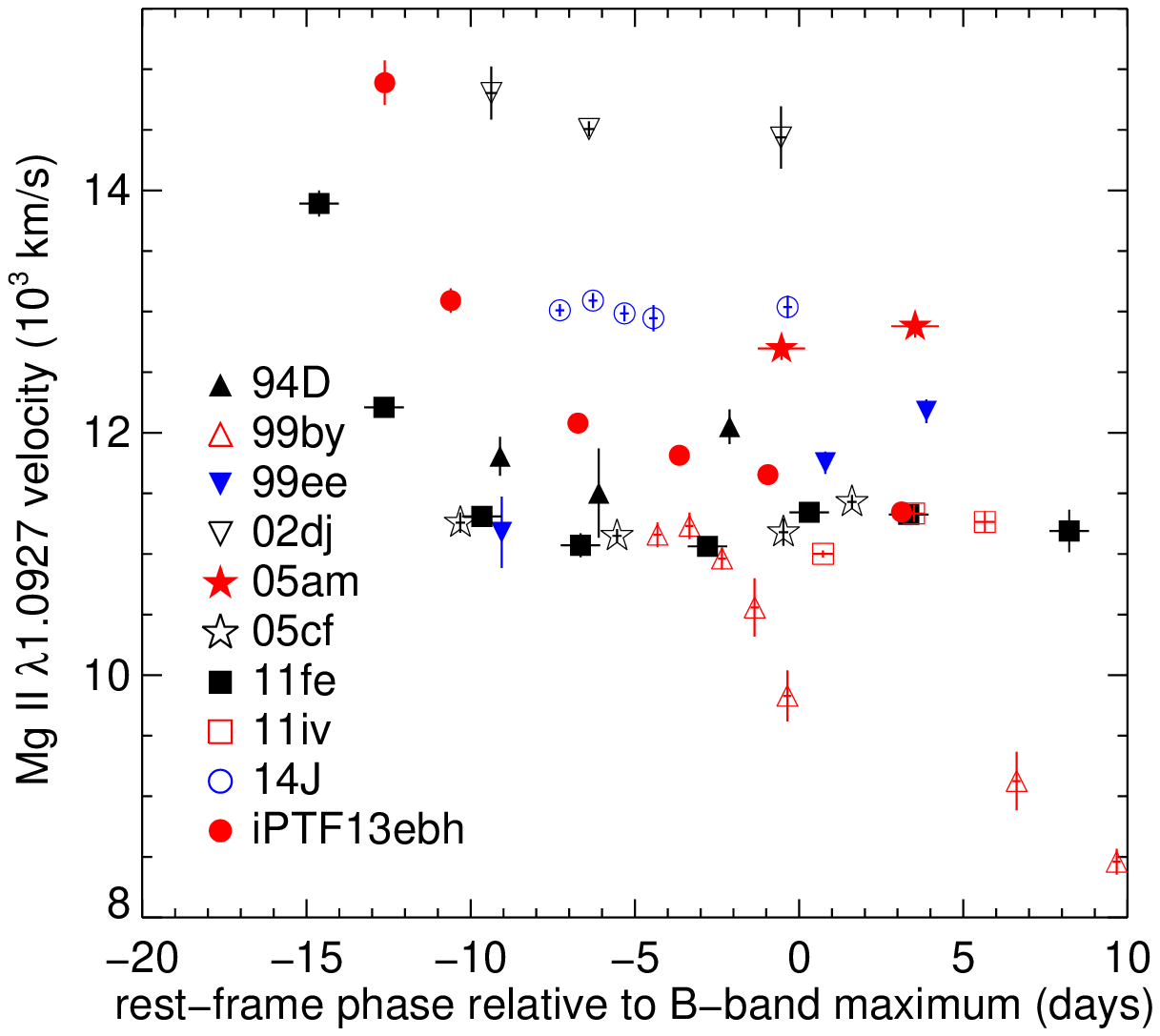}
\caption{The time evolution of the \mgii\ \lam1.0972 \um\ velocity.
  Various symbols represent different objects as noted on the plot.
  Blue, black, and red symbols represent SNe~Ia with optical
  light-curve decline rates in the range \dm<1.0, 1.0<\dm<1.6,
  \dm>1.6, respectively.  The \mgii\ \lam1.0972 \um\ velocity of
  iPTF13ebh decreases monotonically, indicating a deeper-reaching
  carbon burning layer than normal SNe~Ia.}
\label{f:vMgII_t}
\end{figure}

We use the same PCA model of the \mgii\ profile presented in
\citet{2013ApJ...766...72H} and the same fitting procedure using the
nonlinear least-squares fitting program of \citet{2009ASPC..411..251M}
to determine the \mgii\ velocities and the uncertainties for iPTF13ebh
and SN~2014J.  These data points are added to Fig.~8 of
\citet{2013ApJ...766...72H} and presented in Fig.~\ref{f:vMgII_t}.
The high-cadence observations of SN~2014J yielded remarkably flat
\mgii\ velocity, similar to that of SN~2011fe.  The \mgii\ velocity of
SN~2014J is moderately high compared to other SNe~Ia.  Note that the
Gaussian fitting method of \citep{2013ApJ...773...53F} gives
systematic differences in velocities of $\sim300$ \kms.  This
illustrates the importance of using a consistent technique within a
comparison sample.

%difference between iPTF13ebh and the rest

At early phase, the \mgii\ velocity of iPTF13ebh shows a rapid decline
similar to SN~2011fe.  In fact, because of the short rise time of
iPTF13ebh, the rapidly declining portions of iPTF13ebh and SN~2011fe
would line up if Fig.~\ref{f:vMgII_t} was plotted in terms of
explosion date.  However, while the velocity evolution of SN~2011fe
flattens to reveal the base of the carbon burning layer, the
\mgii\ velocity of iPTF13ebh continues to decline.  The \mgii\ line
never reaches a minimum velocity while visible, indicating a deeper
carbon burning layer than normal SNe~Ia.  In this respect, iPTF13ebh
is similar to SN~1999by.

%%%%%%%%%%%%%%%%
%% Transition %%
%%%%%%%%%%%%%%%%

\section{Transitional supernovae}
\label{s:trans}

The term ``91bg-like,'' in the strictest sense, refers to
spectroscopically peculiar SNe~Ia that exhibit strong \tiii\ features
in their maximum-light optical spectra, like SNe~1991bg
\citep{1992AJ....104.1543F, 1993AJ....106.2383B, 1993AJ....105..301L}
and 1999by \citep{2004ApJ...613.1120G}.  These objects are found to be
subluminous, have exceptionally fast-declining light curves, and are
thought to originate from old population progenitors
\citep{2001ApJ...554L.193H}.  Subsequent studies revealed an apparent
bimodality among fast-declining events when their NIR photometric
properties are examined \citep[e.g.,][]{2009AJ....138.1584K,
  2010AJ....139..120F, 2012PASA...29..434P}.  Events with their
primary NIR maxima peaking after the $B$-band maximum are subluminous
in their absolute magnitudes in all bands compared to normal SNe~Ia,
even after taking the width-luminosity relation into account.  These
subluminous supernovae also tend to lack or have very weak secondary
NIR maxima.  The terms ``91bg-like'' and ``subluminous'' are
spectroscopic and photometric definitions, respectively.  So far, all
SNe~Ia with late-peaking NIR maxima have been found to have the
spectroscopic peculiarity of strong \tiii, although, not all SNe~Ia
with \tiii\ signatures have late-peaking NIR maxima, as we will point
out next.

%transitional objects

When defining ``transitional'' events between the fastest-declining
normal SN~Ia population and the 91bg-like/subluminous events, we chose
the timing of the NIR primary maximum rather than the presence of
\tiii\ as the defining feature.  A ``transitional'' event is then
defined as a fast-declining SN~Ia with its NIR ($iYJHK$) primary
maximum peaking before its $B$-band maximum.  The case of SN~1986G
\citep{1987PASP...99..592P} is the main driver behind this choice of
definition.  For SN~1986G, \tiii\ is present in its spectra, like
SN~1991bg, but its NIR absolute magnitudes put it with the normal
SN~Ia group.  It is ``91bg-like'' spectroscopically, but not
``subluminous'' photometrically.  Its NIR maxima also peaks before $B$
maximum.  In our definition, SN~1986G is categorized as
``transitional'' with the normal SNe~Ia and not with the
``subluminous'' group.  More recently, SN~2003gs is also categorized
as ``transitional'' and shows strong
\tiii\ \citep{2009AJ....138.1584K}.  Readers should keep in mind that
``91bg-like'' and ``subluminous'' are, strictly speaking, not the same
group.  However, since our discussions have been focused on
spectroscopic properties, and our only NIR spectroscopic reference in
the 91bg-like and subluminous groups is SN~1999by, we use the two
terms interchangeably.

%iPTF13ebh

From Fig.~\ref{f:phot}, the primary maxima in $iYJH$ light curves of
iPTF13ebh clearly peak before its $B$ maximum.  And, unlike SNe~1986G
and 1991bg, iPTF13ebh shows no evidence of \tiii\ in its optical
spectra (Section~\ref{s:spec:opt}).  This is clearly a transitional
event and has one of the fastest light-curve decline rates in this
class.  For comparison, the decline rates of some of the more notable
transitional events are listed in Table~\ref{t:trans}.

\begin{table}
\caption{Optical light-curve parameters and the presence of \tiii of
  notable ``transitional'' events}
\label{t:trans}
\centering
{\footnotesize
\begin{tabular}{lcccl}
\hline\hline
Name & \dm\ & \sbv\ & \tiii & Data\\
\hline\\[-2ex]
SN~1986G  & $1.62\pm0.02$ & $0.65\pm0.05$ & 91bg-like & (a)\\
SN~2003gs & $1.79\pm0.04$ & $0.46\pm0.03$ & 91bg-like & (b)\\
SN~2003hv & $1.48\pm0.04$ & $0.76\pm0.07$ & No \tiii  & (c)\\
SN~2004eo & $1.42\pm0.03$ & $0.83\pm0.03$ & No \tiii  & CSP; (d)\\
SN~2005am & $1.53\pm0.02$ & $0.75\pm0.02$ & No \tiii  & CSP; (d)\\
SN~2007on & $1.89\pm0.01$ & $0.55\pm0.02$ & No \tiii  & CSP; (e)\\
SN~2009an & $1.64\pm0.04$ & $0.86\pm0.06$ & No \tiii  & (f)\\
SN~2011iv & $1.76\pm0.02$ & $0.63\pm0.01$ & No \tiii  & CSP; (g)\\
SN~2012ht & $1.30\pm0.04$ & $0.86\pm0.03$ & No \tiii  & CSP\\
iPTF13ebh & $1.79\pm0.01$ & $0.63\pm0.02$ & No \tiii  & CSP\\
\hline
\end{tabular}
} \tablefoot{The \dm\ and \sbv\ values are measured using the data
  from the quoted sources.
  \tablefoottext{a}{\citet{1987PASP...99..592P}}
  \tablefoottext{b}{\citet{2009AJ....138.1584K}}
  \tablefoottext{c}{\citet{2009A&A...505..265L}}
  \tablefoottext{d}{\citet{2010AJ....139..519C}}
  \tablefoottext{e}{\citet{2011AJ....142..156S}}
  \tablefoottext{f}{\citet{2013MNRAS.430..869S}}
  \tablefoottext{g}{Gall et al. (in preparation)} }
\end{table}

%sBV

While the width-luminosity relation \citep{1993ApJ...413L.105P} holds
well for a large range of luminosities and \dm, there is a notable
break at \dm$\gtrsim$1.6, where SNe~Ia become faint very quickly with
increased \dm.  Possibly related to this is the fact that \dm\ fails
to properly classify the fastest-declining objects.  The reason for
this is that the onset of the linear decline in the $B$-band light
curve occurs earlier than 15 days for the fastest decliners
\citep{2010AJ....139..120F, 2011AJ....141...19B}.  This leads to
poorly-behaved light-curve templates and a seeming bimodality in the
photometric properties of SNe~Ia with \dm.  Use of \dm\ is therefore
problematic when dealing with transitional objects.
\citet{2014ApJ...789...32B} introduced a new decline-rate parameter,
\sbv, in an attempt to improve the treatment of the fast-declining
objects.  Using \sbv, the fastest declining events appear less as a
distinct population and more as a continuous tail end of normal
SNe~Ia.

%Gonzalez fits

\begin{figure}
\centering
\includegraphics[width=0.47\textwidth,clip=true]{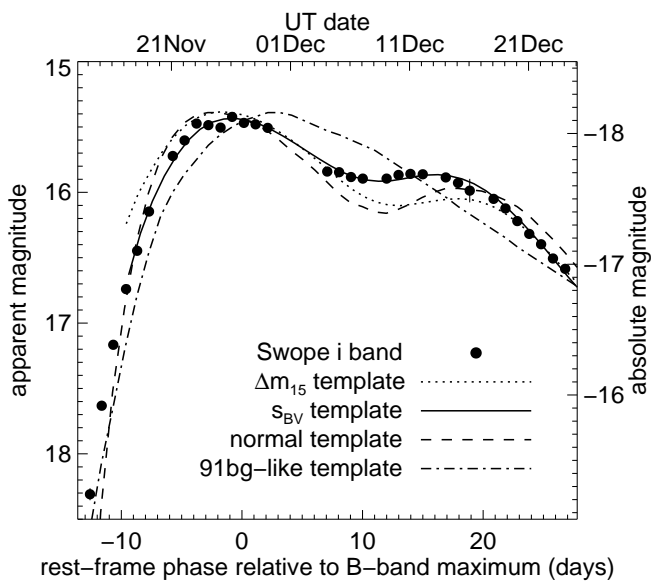}
\caption{Comparison of the Swope $i$-band light curve of iPTF13ebh to
  the \texttt{SNooPy} and \texttt{SiFTO} templates.  The best
  \texttt{SNooPy} fits using \dm\ and \sbv\ $i$-band templates are
  plotted as dotted and solid curves, respectively.  The best
  \texttt{SiFTO} fits with ``normal'' and ``91bg-like'' templates are
  plotted as dash and dot-dash curves, respectively.}
\label{f:snpy}
\end{figure}

In Fig.~\ref{f:snpy}, the Swope $i$-band light curve of iPTF13ebh is
plotted in comparison with the ``normal'' and ``91bg-like'' fits, as
described by \citet{2014ApJ...795..142G}.  The fits are done with the
light-curve fitter \texttt{SiFTO} \citep{2008ApJ...681..482C}, but
with two sets of light-curve templates for comparison: one of normal
SNe~Ia and the other constructed from 91bg-like objects.  The fitter
stretches the time axis to fit the iPTF13ebh data from all of the
optical $uBVgri$ bands.  The result shows poor fits in the $i$ band
for both ``normal'' and ``91bg-like'' templates.  iPTF13ebh has an
$i$-band light curve morphology that bridges the two classes, with
transitional properties, such as the timing of the primary and
secondary peaks and the size of the secondary peak.

%SNooPy fits

Fig.~\ref{f:snpy} also shows comparison \texttt{SNooPy} fits using
\dm\ and \sbv\ templates.  The fits also include all optical $uBVgri$
bands.  Unlike \texttt{SiFTO} fits, \texttt{SNooPy} takes into account
the changing shape of the light curves in the redder filter bands for
supernovae with different peak luminosities.  The \dm\ template fit
yields an $i$-band light curve with a primary maximum that peaks too
early and a secondary maximum that peaks too late, which is
characteristic of a brighter supernova than the $i$-band light curve
suggests and is much like the ``normal'' \texttt{SiFTO} fit described
above.  The \sbv\ template, on the other hand, provides an excellent
fit to the observed light curve.  The effect would be even more
pronounced in the NIR $YJH$ bands.  We chose the $i$-band light curve
for illustration, as it has much better coverage than the NIR bands
(Fig.~\ref{f:phot}).

%conclusion

The photometric properties of iPTF13ebh place it with the normal
SNe~Ia, but many of its NIR spectroscopic properties, such as the
strong \ci, the profile shape of the $H$-band feature, and the
\mgii\ velocity evolution, are similar to those of 91bg-like
SN~1999by.  \texttt{SNooPy} with the \sbv\ parameterization works well
for iPTF13ebh and other transitional objects.  This suggests a
continuous range of properties across the transition.  However, the
\sbv\ parameter does not predict the presence of \tiii\ (nor does \dm,
see Table~\ref{t:trans}).  It is unclear whether normal and
subluminous/91bg-like belong in distinct groups or are objects with a
common explosion mechanism and a continuous range of observed
properties, but iPTF13ebh provides a connection between these
subtypes.

%%%%%%%%%%%%%%%%%
%% Conclusions %%
%%%%%%%%%%%%%%%%%

\section{Conclusions}
\label{s:conc}

iPTF13ebh is one of the fastest-declining and best-observed
``transitional'' objects between normal-bright and
subluminous/91bg-like events.  Overall, iPTF13ebh has properties that
are distinct from both normal and subluminous classes.  We
summarize the main findings as follows:

\begin{enumerate}
\item The NIR $iYJH$ light curves of iPTF13ebh peak before its $B$
  maximum, and its extinction-corrected absolute magnitudes obey the
  width-luminosity relation.
\item Several \ci\ lines are detected in the early NIR spectra of
  iPTF13ebh.  The \ci\ \lam1.0693 \um\ line at 2.3 days past explosion
  ($-12.8$ days relative to $B$ maximum) is the strongest ever
  observed in a SN~Ia.
\item The profile of the \ci\ \lam1.0693 \um\ line at first resembles
  that of SN~1999by, a subluminous/91bg-like object.  However, while
  the strong \ci\ line persists until maximum light in SN~1999by, the
  \ci\ line of iPTF13ebh quickly weakens and takes on a profile
  similar to that of a normal SN~Ia.
\item The NIR \ci\ \lam1.0693 \um\ line of iPTF13ebh is much stronger
  than its optical \cii\ \lam0.6580 \um\ line.  This property is
  general among SNe~Ia with \ci\ detections, and contradicts the
  prediction of stronger \cii\ lines by \citet{2008ApJ...677..448T}.
\item Unlike normal SNe~Ia, iPTF13ebh shows flat \ci\ velocity
  evolution, indicating that the base of the unburned layer is already
  reached at the earliest phases.
\item The $H$-band break ratio of iPTF13ebh fits well in the peak
  ratio versus \dm\ relation previously found by
  \citet{2013ApJ...766...72H}.  The strong correlation is consistent
  with the Chandrasekhar mass delayed detonation scenario, while the
  correlation is expected to be weak in dynamical mergers.
\item By late time, the profile shape of the $H$-band complex
  resembles that of SN~1999by.  The composition and density of the
  inner core of iPTF13ebh is therefore similar to that of 91bg-like
  events.
\item Unlike in normal SNe~Ia, the \mgii\ velocity of iPTF13ebh
  monotonically decreases with time indicating a deeper reaching
  carbon burning layer than normal SNe~Ia.  This behavior is similar
  to that of SN~1999by.
\item There is no evidence of a detached high-velocity component for
  the \caii\ NIR triplet of iPTF13ebh, as early as 3.2 days past
  explosion, providing a stringent constraint on when this
  high-velocity feature could have formed, if it ever existed.
\item iPTF13ebh has a substantial difference between
  the explosion times inferred from the early-time light curve and the
  velocity evolution of the \siii\ \lam0.6355 \um\ line, implying a
  long dark phase of $\sim 4$ days.

\end{enumerate}

The data set presented here is unique, as the follow up campaign began
shortly after explosion and spanned a wide wavelength coverage from
the UV to the NIR.  iPTF13ebh is also the only ``transitional'' SN~Ia
with extensive NIR spectroscopic coverage.  The photometric properties
of iPTF13ebh suggest that it is at the fast-declining end of normal
SNe~Ia.  Yet, the NIR spectra reveal many similarities to the
91bg-like SN~1999by.  The rare glimpse of the NIR spectroscopic
properties of a transitional event provides a connection between
normal and the peculiar subluminous/91bg-like events, perhaps
indicating a common explosion mechanism.

%%%%%%%%%%%%%%%%%%%%%
%% Acknowledgments %%
%%%%%%%%%%%%%%%%%%%%%

\begin{acknowledgements}

This paper is based upon work supported by the National Science
Foundation under Grant No. AST-1008343.  M.~S., E.~Y.~H., C.~C, and
C.~G. acknowledge the generous support provided by the Danish Agency
for Science and Technology and Innovation through a Sapere Aude Level
2 grant.  S.~G. acknowledges support from CONICYT through FONDECYT
grant 3130680 and from the Ministry of Economy, Development, and
Tourism's Millennium Science Initiative through grant IC12009, awarded
to The Millennium Institute of Astrophysics, MAS.  LANL participation
in iPTF is supported by the US Department of Energy as part of the
Laboratory Directed Research and Development program.  The bulk of the
data presented here was obtained with the 1-m Swope, 2.5-m du Pont and
the 6.5-m Magellan Telescopes at the Las Campanas Observatory.  This
work also relies on data obtained at the Gemini Observatory, under the
long-term program GN-2013B-Q-76.  The Gemini Observatory is operated
by the Association of Universities for Research in Astronomy, Inc.,
under a cooperative agreement with the NSF on behalf of the Gemini
partnership: the National Science Foundation (United States), the
National Research Council (Canada), CONICYT (Chile), the Australian
Research Council (Australia), Minist\'{e}rio da Ci\^{e}ncia,
Tecnologia e Inova\c{c}\~{a}o (Brazil) and Ministerio de Ciencia,
Tecnolog\'{i}a e Innovaci\'{o}n Productiva (Argentina).  The authors
would like to recognize the very significant cultural role and
reverence that the summit of Mauna Kea has within the indigenous
community of Hawaii.  We are grateful for our opportunity to conduct
observations from this mountain.  We have also made use of the Nordic
Optical Telescope, which is operated by the Nordic Optical Telescope
Scientific Association at the Observatorio del Roque de los Muchachos,
La Palma, Spain, of the Instituto de Astrofisica de Canarias.  The
William Herschel Telescope and its override programme are operated on
the island of La Palma by the Isaac Newton Group in the Spanish
Observatorio del Roque de los Muchachos of the Instituto de
Astrof\'isica de Canarias.  This research used resources from the
National Energy Research Scientific Computing Center (NERSC), which is
supported by the Office of Science of the U.S. Department of Energy
under Contract No. DE-AC02-05CH11231.  We have also made use of the
NASA/IPAC Extragalactic Database (NED) which is operated by the Jet
Propulsion Laboratory, California Institute of Technology, under
contract with the National Aeronautics and Space Administration.

\end{acknowledgements}

%%%%%%%%%%%%%%%%%%
%% Bibliography %%
%%%%%%%%%%%%%%%%%%

\bibliographystyle{aa}
\bibliography{bibliography}

\end{document}